\documentclass[journal,10pt,twocolumn]{IEEEtran}
\usepackage{amsmath,amsfonts,amssymb}
\usepackage{algorithmic}
\usepackage{array}
\usepackage{tabularx}
\usepackage[caption=false,font=normalsize,labelfont=sf,textfont=sf]{subfig}
\usepackage{textcomp,graphics}
\usepackage{stfloats}
\usepackage{url}
\usepackage{verbatim}
\usepackage{graphicx}
\usepackage{epsfig}
\usepackage{xcolor}
\usepackage[colorlinks=true,linkcolor=red,citecolor=blue,urlcolor=blue,]{hyperref}
\usepackage[capitalize]{cleveref}
\newtheorem{lemma}{Lemma}
\newtheorem{theorem}{Theorem}
\newtheorem{remark}{Remark}
\newtheorem{corollary}{Corollary}
\newtheorem{definition}{Definition}
\newcounter{MYtempeqncnt}
\def\BibTeX{{\rm B\kern-.05em{\sc i\kern-.025em b}\kern-.08em
		T\kern-.1667em\lower.7ex\hbox{E}\kern-.125emX}}

\begin{document}
	
	\title{Fluid Antenna Multiple Access Assisted\\Integrated Data and Energy Transfer: Outage and Multiplexing Gain Analysis}
	
	\author{Xiao Lin, Yizhe Zhao,~\IEEEmembership{Member,~IEEE,} Halvin Yang, Jie Hu,~\IEEEmembership{Senior Member,~IEEE}, Kai-Kit Wong,~\IEEEmembership{Fellow,~IEEE}
		\vspace{-9mm}
		% <-this % stops a space
		\thanks{Xiao Lin, Yizhe Zhao and Jie Hu are with the School of Information and Communication Engineering, University of Electronic Science and Technology of China, Chengdu 611731, China. (e-mail: xiaolin@std.uestc.edu.cn; yzzhao@uestc.edu.cn; hujie@uestc.edu.cn).}
		\thanks{Halvin Yang and Kai-Kit Wong are with the Department of Electronic and Electrical Engineering, University College London, Torrington Place, WC1E 7JE, United Kingdom (e-mail:uceehhy@ucl.ac.uk; kai-kit.wong@ucl.ac.uk). K. K. Wong is also affiliated with Yonsei Frontier Lab, Yonsei University, Seoul, Korea.}
	}
	
	\maketitle
	
	\begin{abstract}
		Fluid antenna multiple access (FAMA) exploits spatial opportunities in wireless channels through port switching to overcome multiuser interference, achieving better performance than traditional fixed MIMO systems. Integrated Data and Energy Transfer (IDET) is capable of providing both wireless data transfer (WDT) and wireless energy transfer (WET) services for low-power devices. This paper investigates an FAMA-assisted IDET system, where a base station (BS) equipped with $N$ fixed antennas provides dedicated IDET services to $N$ user equipments (UEs). Each UE is equipped with a single fluid antenna, while the power splitting (PS) approach is conceived for coordinating WDT and WET.  Under the Rayleigh channel model, we derive both exact expressions and approximate closed forms for the outage probabilities of WDT and WET, where the fluid antenna (FA) at each UE selects the optimal port to maximize either the signal-to-interference-plus-noise ratio (SINR) or the energy harvesting power (EHP). The IDET outage probabilities are defined and subsequently derived and approximated into closed-forms. Further, multiplexing gains of the proposed system are defined and analyzed to evaluate the performance.  Further, we analyze the IDET outage probabilities and multiplexing gains of the proposed system. Additionally, to provide a more general analysis, we extend our analytical framework to the Rician channel model. Numerical results validate the theoretical analysis while also illustrating that the trade-off is achieved between WDT and WET performance by exploiting different port selection strategies and numbers of UEs. 
	\end{abstract}
	
	
	\section{Introduction}
	\subsection{Background}
	The rapid advances in mobile communications technology mean that massive low-power devices are swarming into the networks for providing a variety of services to smart factories and smart cities, etc.~\cite{Ontheroad}. It is posing formidable challenges to energy supplements for such devices, as manually charging or replacing batteries entails additional human costs, a main bottleneck in many applications \cite{Distributed}. Integrated data and energy transfer (IDET) has been envisioned to address the battery life issues for low-power devices in next-generation wireless networks \cite{L.R.Varshney}. Usually, devices have both data and energy requirements, and thus wireless data transfer (WDT) and wireless energy transfer (WET) should be coordinated at the transceiver, which yields the concept of IDET \cite{Rzhang,I.Krikidis}.
	
	Traditional multiple input multiple output (MIMO) technology enhances the performance of IDET systems in congested networks using spatial multiplexing, diversity coding, and beamforming techniques \cite{9720149,7009979}. However, MIMO is usually associated with increased energy consumption and hardware costs, with the need of multiple antennas and radio frequency (RF) chains, which is often impractical for low-power devices under limited hardware resources. To address these challenges, fluid antenna (FA) has emerged in recent years as a promising solution for future wireless communication systems \cite{Wong-2022fcn,wong2023fluid}. An FA system (FAS) represents the general concept of shape-flexible and position-flexible antenna technologies, which may be implemented adopting surface wave technology \cite{Shen-tap_submit2024} and reconfigurable pixels \cite{Zhang-pFAS2024}. The latter can achieve near-zero-delay switching of antenna position for communications. There is also a branch of FAS known as `movable antenna' \cite{Zhu-June 2024}, which  can be implemented using the motor slide-based architecture, micro-electro-mechanical systems (MEMS) integrated antennas, deployable antennas, or other structural designs \cite{Zhu-June 2024-TWC}. Specifically, Zhu \emph{et al.} investigated movable antennas for enhancing multiple access channels and beamforming in \cite{Zhu-July 2024-TWC} and \cite{Zhu-Dec 2024-CL}, respectively. It is anticipated that FAS has great potential in IDET systems.
	
	Regarding IDET systems, the received signals often suffer from serious power attenuation in the transmission path, which will degrade the IDET efficiency, especially for WET, whose activating power threshold is much higher than that of WDT \cite{9257429}. Hence, to glean sufficient energy for functioning reliably, FAS can be useful in providing higher IDET efficiency with a much smaller hardware size compared to traditional fixed-position MIMO technology. Specifically, the spatial diversity gain from FA can significantly enhance the capability of low-power devices to achieve better performance in both WDT and WET, which is more suitable for an IDET system.
	
	\vspace{-2mm}
	\subsection{Related Work}
	In most research results, WDT and WET are coordinated by one of the two common approaches, namely, time switching (TS) and power splitting (PS) \cite{Rzhang,I.Krikidis}. In the TS approach, a time switch is applied at the receiver to allow user equipments (UEs) to perform information decoding (ID) or energy harvesting (EH) in different time slots. By contrast, in the PS approach, a power splitter at the UE is utilized to divide the received signal into two parts, one for ID and another for EH. Results for the TS approach and the PS approach have been reported in various wireless systems, such as broadcast channel (BC) \cite{HLee},  cooperative non-orthogonal multiple-access (NOMA) network \cite{R.R.KurupandA.V.Babu}, interference channels (IFCs) \cite{Exploiting,Beamforming,A.A.Nasir} and etc. Specifically, Lee \emph{et al.} considered a two-user IDET multiple-input single-output  (MISO) BC with a new joint TS approach \cite{HLee}, and Kurup \emph{et al.} then investigated an adaptive power allocation scheme for the IDET-enabled cooperative NOMA network under the TS approach \cite{R.R.KurupandA.V.Babu}. Also, \cite{Exploiting} introduced a novel precoding design, while \cite{Beamforming} proposed beamforming designs, ensuring both Quality-of-Service (QoS) and the EH constraints in the MISO channel. Furthermore, \cite{A.A.Nasir} addressed the joint design of transmit beamforming  with either PS ratios or transmit TS ratios in a multicell network. Moreover, Zhao \emph{et al.} proposed a novel time index modulation assisted IDET system, where data is delivered not only by the phase-amplitude modulation but also by the multiple activated time indices for either WDT or WET \cite{Time,TIM}. 
	
	On the other hand, FAS research is comparatively a newer endeavor. The use of FAS for wireless communications was first proposed by Wong {\em et al.} in 2020 \cite{PerformanceLimits,FAsystem}. Subsequently, results have been presented on the performance of FAS for single-user communication, focusing on the outage probability and diversity gain \cite{AnalyticalApproximation,NewInsights,OntheDiversity,Vega-2023}. A two-stage approximation incorporating more parameters into the channel model was proposed in \cite{AnalyticalApproximation}, offering a more accurate characterization of the performance for FAS. In \cite{NewInsights}, the authors simplified the outage probability at high signal-to-noise ratio (SNR) and obtained the diversity gain of FAS. By combining space-time rotations with code diversity, Psomas \textit{et al.} designed space-time coded modulations that attain promising performance over block-fading channels \cite{OntheDiversity}. Later in \cite{Vega-2023} and \cite{Vega-2023-2}, the authors used an asymptotic matching method to analyze the performance of FAS under Nakagami fading channels. Other work also studied the performance of FAS under $\alpha$-$\mu$ fading channels \cite{Alvim-2023} and considered the continuous FAS where the change of position can be infinitely fine \cite{Psomas-dec2023}. Most recently, the MIMO-FAS setup where both ends have FAS with multiple activated ports was also investigated in \cite{New-twc2023}.
	
	FAS is more than just a new degree of freedom for design and optimization. In \cite{fFAMA,Massive,sfama}, it was shown that position flexibility enabled by FAS can simplify multiple access and allow spectrum sharing on the same physical channel without precoding and without interference cancellation. The approach in \cite{fFAMA,Massive} is dubbed as fast fluid antenna multiple access (FAMA) which switches antenna position at each UE on a per-symbol basis to mitigate inter-user interference. Fast FAMA is powerful but is not known to be practically achievable. By contrast, the slow FAMA scheme in \cite{sfama} is more practical as it only requires each UE to switch its FA port once in each channel coherence time. In \cite{Revisiting}, the performance of the two-user slow FAMA channel was accurately studied while \cite{yang} adopted the Gauss-Laguerre and Gauss-Hermite quadrature to conduct the performance analysis applicable to any number of UEs. Deep learning has also been utilized to reduce the burden of signal-to-interference-plus-noise ratio (SINR) observation at the ports \cite{N.Waqar}. Later in \cite{Y.Chen}, the authors developed an optimal stochastic control algorithm based on mean-field game theory to address the power control problem in the slow FAMA system. Advanced system models involving FAS such as integrated sensing and communication (ISAC) \cite{ISAC_FAS}, dirty multiple access channels \cite{F.RostamiGhadi} and NOMA \cite{W.K.New} have also been considered. 
	
	\subsection{Motivation and Contributions}  
	However, there are few works studying FAMA-assisted IDET. In our previous work \cite{lin}, we analyzed the outage probabilities for WDT and WET in a slow FAMA-assisted IDET system, where the TS approach was conceived. It was shown that FAMA-assisted IDET outperforms traditional MIMO-assisted IDET in both WDT and WET performance. In general, existing results showed that the PS approach always achieves better IDET performance than the TS approach, which requires accurate synchronization between transmitters and receivers. To the best of our knowledge, integrating FAS into IDET systems with the PS approach remains unexplored. In the PS approach, port selection strategy simultaneously affects both the WDT and WET, making performance analysis more challenging compared to that of the TS approach, where WDT and WET can be segregated into different time slots.
	
	This paper investigates FAMA-assisted IDET systems using the PS approach to coordinate WDT and WET at the UEs. We assume slow FAMA, where each UE selects the optimal port for maximum SINR or energy harvesting power (EHP). We focus on outage probabilities and multiplexing gains for WDT and WET based on these strategies. Additionally, we define and analyze special and general IDET outage probabilities and multiplexing gains, considering both WDT and WET simultaneously. Specifically, our contributions are summarized as follows:
	\begin{itemize}
		\item This paper investigates the FAMA-assisted IDET system using PS to coordinate WDT and WET, where each UE is equipped with a single $K$-port FA. Each UE dynamically  selects the optimal receive port according to various WDT and WET requirements.
		\item Under the Rayleigh channel model, exact and approximate closed-form expressions are derived for the outage probabilities of WDT and WET, considering two port selection strategies: a WDT-oriented strategy maximizing SINR and a WET-oriented strategy maximizing EHP. Additionally, the multiplexing gains for WDT and WET are defined and analyzed.
		\item Both the special and general IDET outage probabilities are analyzed without considering a specific port selection strategy, thereby offering a more comprehensive view of the proposed system. Then the IDET outage probabilities are approximated into closed-forms, and the corresponding IDET multiplexing gains are defined and analyzed.
		\item  The exact outage probabilities for WDT and WET are derived under the Rician channel model to provide a more general analytical framework.
		\item Simulation results validate the theoretical analysis and offer some novel insights for the design of FAMA-assisted IDET systems.
	\end{itemize}
	
	The rest of this paper is organized as follows. Section \ref{systemmodel} introduces the system model of the FAMA-assisted IDET system. In
	Section \ref{section3} and Section \ref{section4}, we analyze the outage probabilities and multiplexing gains of the proposed system under the Rayleigh channel model. Section \ref{section7} extends the analysis to the Rician channel model. After providing simulation results in Section \ref{section5}, we conclude this paper.
	
	\section{System Model}\label{systemmodel}
\begin{figure}[t]
		\centering
		\includegraphics[width=3in]{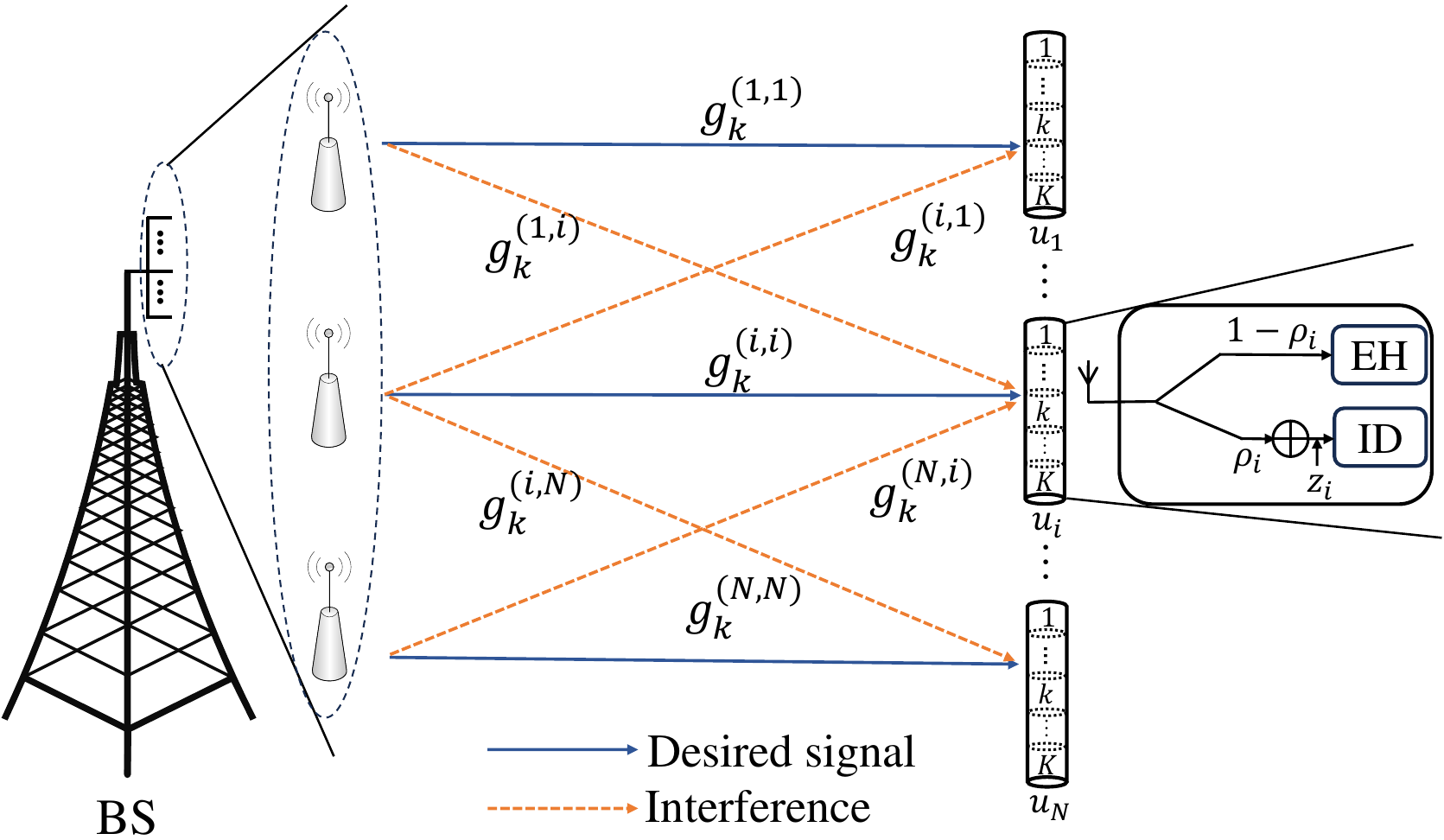}
		\caption{An FAMA-assisted IDET system.}\label{model}
	\end{figure}

	Consider a downlink FAMA-assisted IDET system consisting of a base station (BS) equipped with $N$ fixed antennas\footnote{Note that the overall performance can be improved if the BS is also equipped with fluid antennas, since the system can benefit from more multiplexing gains. However, the theoretical performance analysis will be much more complex and difficult, which will be considered in our future works.} and $N$ UEs. The BS antennas are distributed far apart so that the channels from all the BS antennas to a specific UE are completely independent. Similarly, we assume that there is no correlation between the channels from each specific BS antenna to any UEs. Each UE is equipped with a $K$-port linear FA having the size of $W\lambda$, where $W$ represents the normalized size of FA and $\lambda$ denotes the wavelength. It is assumed that each BS antenna is dedicated to communicating with its corresponding UE\footnote{The digital beamforming is not considered at the BS side since it requires additional channel estimation operations to obtain accurate channel state information (CSI). Moreover, beamforming would necessitate extra precoding and power allocation among antennas, thereby increasing the complexity of signal processing.}, \textit{i.e.}, the $i$-th BS antenna communicates with UE $i$\footnote{In our paper, the BS antennas are assumed to be sufficiently spaced apart, ensuring that the channels between each BS antenna and the UEs remain statistically independent.  Thus, reallocating antenna-UE associations does not change the channel statistics, making antenna-UE association optimization unnecessary in this case.}, and each UE uses the PS approach to coordinate WDT and WET, as shown in Fig.~\ref{model}. The UEs may also receive the interference signals from other BS antennas, which may degrade the WDT performance but provide more signal sources for WET.
	
	\subsection{Rayleigh  channel model}
	It is assumed that the channel is modeled in the rich scattering environment, thus there is no line-of-sight (LoS) component. The channels are considered to be correlated since they can be close to each other.  Similar to \cite{sfama}, we characterize the channel between the $m$-th BS antenna and the $k$-th port of UE $i$ through a single correlation parameter $\mu$ as
	\begin{multline}\label{eqgk}
		g_k^{(m,i)}=\left(\sqrt{1-\mu^2}x_k^{(m,i)}+\mu x_0^{(m,i)}\right)\\
		+j\left(\sqrt{1-\mu^2}y_k^{(m,i)}+\mu y_0^{(m,i)}\right),
	\end{multline}
	where $x_0^{(m,i)},\dots,x_K^{(m,i)}$ and $y_0^{(m,i)},\dots,y_K^{(m,i)}$ are independent Gaussian random variables with zero mean and variance of $1$. Note that the correlation among different ports is realized by invoking the same variables $x_0^{(m,i)} , y_0^{(m,i)}$. The choice of $\mu$  aims to simulate the average squared spatial correlation of an actual fluid antenna structure, which is given by \cite{sfama}
	\begin{equation}\label{mucalculate}
		\mu=\sqrt{2}\sqrt{_{1}F_2\left(\frac{1}{2};1,\frac{3}{2};-\pi^2W^2\right)-\frac{J_1(2\pi W)}{2\pi W}},
	\end{equation}
	where $_{a}F_b(\cdot;\cdot;\cdot)$ denotes the generalized hypergeometric function and $J_1(\cdot)$ is the first-order Bessel function of the first kind.

	\subsection{IDET with PS approach}	
	UE $i$ applies the PS approach to coordinate WDT and WET from the received signal. Specifically, the received signal at each UE is split for the WDT and WET by a power spitter, which divides $\rho_i $ $(0 \leq \rho_i \leq 1)$ portion of the signal power for WDT, and the remaining $1-\rho_i$ portion of power for WET. Each UE is assumed to have its own requirements for the SINR of WDT and the EHP of WET. Then, the received signal for WDT at the $k$-th port of UE $i$ is expressed as
	\begin{multline}\label{ykWDT}
		y_{k,i}^{\text{WDT}}=\underbrace{\sqrt{\frac{\rho_iP_{i,i}}{d_{i,i}^\beta}}g_k^{(i,i)}s_{i}}_{\text{desired signal}}+
		\underbrace{\sum_{m\neq i}\sqrt{\frac{\rho_iP_{m,i}}{d_{m,i}^\beta}}g_k^{(m,i)}s_{m}}_{\text{interference}}\\+\underbrace{\sqrt{\rho_i}\eta_k^i+z_i}_{\text{noise}},
	\end{multline}
	where $s_{i}$ denote the transmit Gaussian signal from the $i$-th antenna for UE $i$, with $P_{i,i}$ and $P_{m,i}$ representing the transmit powers from the $i$-th BS antenna to UE $i$ and from the 
	$m$-th BS antenna to UE $i$, respectively.  $d_{i,i}$ and $d_{m,i}$ correspond to the distances  from the 
	$i$-th BS antenna to UE  $i$ and 
	$m$-th BS antenna to UE $i$, respectively. $\beta$ is the path loss exponent. Moreover, $\eta_k^i$ is the complex additive white Gaussian noise (AWGN) with zero mean and variance of $\sigma^2_\eta$ at the $k$-th port of UE $i$, and $z_i$ is the passband-to-baseband noise of UE $i$, which also follows the complex Gaussian distribution with zero mean and variance of $\sigma^2_c$.  Accordingly, the received SINR at $k$-th port of UE $i$ is given by
	\begin{equation}\label{SINR}
		\begin{aligned}
			\text{SINR}_k^i&\hspace{.5mm}=\frac{\rho_i\frac{P_{i,i}}{d_{i,i}^\beta}\left|g_k^{(i,i)}\right|^2 \mathbb{E}[\Vert s_{i} \Vert ^2]}{\rho_i\sum_{m\neq i}^{N}\frac{P_{m,i}}{d_{m,i}^\beta}\left|g_k^{(m,i)}\right|^2 \mathbb{E}[\Vert s_{m} \Vert ^2]+(\rho_i\sigma_{\eta}^2+\sigma_{c}^2)}\\
			&\overset{(a)}{\approx}\frac{\rho_i\frac{P_{i,i}}{d_{i,i}^\beta}\left|g_k^{(i,i)}\right|^2 }{\rho_i\sum_{m\neq i}^{N}\frac{P_{m,i}}{d_{m,i}^\beta}\left|g_k^{(m,i)}\right|^2 }\overset{(b)}{=}\frac{\left|g_k^{(i,i)}\right|^2 }{\sum_{m\neq i}^{N}\left|g_k^{(m,i)}\right|^2},
		\end{aligned}
	\end{equation}
	where $(a)$ assumes that the interference power is much greater than the noise power, and the SINR is reduced to the signal-to-interference-ratio (SIR), $(b)$ assumes the case that all the BS antennas have identical transmit power $P$\footnote{Since we focus on the performance analysis of the FAMA-assisted IDET system but not the power allocation among BS antennas, it is assumed that all the BS antennas have identical transmit power $P$ for analytical tractability.} and all the distances from the BS antennas to UE $i$ is the same ($d_{m,i}=d, \forall i$)\footnote{Since the distance between the BS and the UEs is much farther than the antenna spacing, the differences among antenna-UE pairs are neglected.}.

	The signal split for WET at UE $i$ by activating the $k$-th port is expressed as
	\begin{equation}\label{ykWET}	y_{k,i}^{\text{WET}}=\sqrt{1-\rho_i}\left(\sum_{m= 1}^N\sqrt{\frac{P_{m,i}}{d_{m,i}^\beta}}g_k^{(m,i)}s_{m}+\eta_k^i\right).
	\end{equation}	
	Then, the EHP for WET at UE $i$ at the $k$-th port is given by\footnote{Since the noise is much smaller than the desired signal and the interference, the energy harvested from noise is negligible.} 
	\begin{equation}\label{Qk}
		Q_{k}^{i}=(1-\rho_i)\frac{P}{d^\beta}\sum_{m=1}^{N}\left|g_k^{(m,i)}\right|^2.
	\end{equation}
	
	\section{WDT and WET Performance Analysis with Different Port Selection Strategies}\label{section3}
	In this section, we will study the outage probabilities for WDT and WET by conceiving different port selection strategies in the FAMA-assisted IDET system. The multiplexing gains for WDT and WET will also be analyzed.
	
	\subsection{Port selection strategy}
	The interference signals at each UE may result in different impacts on the performance of WDT and WET. For instance, a stronger interference may cause ID failure but might increase the EHP for WET. Consequently, the FA can adjust the port selection strategy to maximize the SINR or the EHP to meet various requirements. On the one hand, if UE $i$ is more eager to achieve better WDT performance, it applies the WDT oriented port selection strategy, where the UE chooses an optimal port that maximizes the SINR in \eqref{SINR}, which is given by\footnote{ In this paper, we assume that the CSI of all the ports is known at each UE for the port selection since it can be  reconstructed  by only requiring a small number of the ports \cite{N.Waqar}.}
	\begin{equation}\label{SINRM}
		k_{\text{SINR}}^{\ast}= \arg\max_k \frac{\left\vert g_k^{(i,i)} \right\vert^2 }{\sum_{m\neq i}^{N}\left\vert g_k^{(m,i)} \right\vert^2 }\triangleq \arg\max_k \frac{X_k}{Y_k},
	\end{equation}
	where
		\begin{equation}\label{X_k}
			X_k=\left(x_k^{(i,i)}+\frac{\mu x_0^{(i,i)}}{\sqrt{1-\mu^2}} \right)^2+
			\left(y_k^{(i,i)}+\frac{\mu y_0^{(i,i)}}{\sqrt{1-\mu^2}} \right)^2,
	\end{equation}
	and
	\begin{equation}\label{Y_k}
	Y_k= \sum_{m=1\atop m\neq i }^{N}\left(x_k^{(m,i)}+\frac{\mu x_0^{(m,i)}}{\sqrt{1-\mu^2}}\right)^2+\left(y_k^{(m,i)}+\frac{\mu y_0^{(m,i)}}{\sqrt{1-\mu^2}} \right)^2.
\end{equation}
	On the other hand, if UE $i$ is more eager to achieve better WET performance, it will apply the WET oriented port selection strategy, where the UE selects an optimal antenna port that maximizes the EHP in \eqref{Qk}, which is given by
	\begin{equation}\label{HPM}
		k_{\text{EHP}}^{\ast}= \arg\max_k \enspace \sum_{m=1}^{N}\left\vert g_k^{(m,i)} \right\vert^2
		\triangleq\arg\max_k \enspace(X_k+Y_k).
	\end{equation}
	
	\subsection{Outage probabilities analysis with WDT oriented port selection strategy}
	It is considered that ID suffers from a failure and WDT experiences an outage if the received SINR at UE $i$ is lower than a certain threshold $\gamma_{\text{th}}$.  Using the port selection strategy in \eqref{SINRM}\footnote{Note that the WDT oriented port selection strategy is the standard port selection approach in FAMA without WET \cite{sfama}.}, we can analyze the outage probabilities of WDT and WET with the WDT oriented strategy, respectively. Firstly, the WDT outage probability is defined as
	\begin{equation}\label{WDToutageexpression}
		\epsilon_{\text{WDT-SINR}}^i=\text{Prob}\left( \max_k\enspace\text{SINR}_k^i =  \frac{X_k}{Y_k} < \gamma_{\text{th}}\right).
	\end{equation}
	
	According to \cite{sfama}, the exact expression for the WDT outage probability $\epsilon^i_{\text{WDT-SINR}}$ with the WDT oriented strategy of UE $i$ is formulated as \eqref{ExactWDTP} (see top of next page), 
	\begin{figure*}[]
		% Store the current equation number.
		%		\setcounter{MYtempeqncnt}{\value{equation}}
		% Set the equation number to one less than the one
		% desired for the first equation here.
		% The value here will have to changed if equations
		% are added or removed prior to the place these
		% equations are referenced in the main text.
		%\setcounter{equation}{11}
		\begin{equation}\label{ExactWDTP}
			\begin{aligned}
				\epsilon_{\text{WDT-SINR} }^i=&\int_{0}^{\infty}\int_{0}^{\infty}\frac{\tilde{r}^{N-2}e^{-\frac{r+\tilde{r}}{2}}}{2^{N}\Gamma(N-1)}\left[  Q_{N-1}\left(\sqrt{\frac{\mu^2\gamma_{\text{th}}\tilde{r}}{\left(1-\mu^2\right)\left(\gamma_{\text{th}}+1\right)}},\sqrt{\frac{\mu^2r}{\left(1-\mu^2\right)\left(\gamma_{\text{th}}+1\right)}}\right)-e^{-\frac{\mu^2\left(\gamma_{\text{th}}\tilde{r}+r\right)}{2\left(1-\mu^2\right)\left(\gamma_{\text{th}}+1\right)}}\right.\\
				&\left.\times\left(\frac{1}{\gamma_{\text{th}}+1}\right)^{N-1} \sum_{k=0}^{N-2}\sum_{j=0}^{N-k-2}\frac{(N-j-k-1)_j}{j!}\left(\frac{r}{\tilde{r}}\right)^{\frac{j+k}{2}}(\gamma_{\text{th}}+1)^k\gamma_{\text{th}}^{\frac{j-k}{2}}I_{j+k}\left(\frac{\mu^2}{1-\mu^2}\frac{\sqrt{\gamma_{\text{th}} r \tilde{r}}}{\gamma_{\text{th}}+1}\right) \right] ^Kdrd\tilde{r}
			\end{aligned}
		\end{equation}
		\hrulefill
	\end{figure*}
	where $(t)_k = t(t+1)\cdots(t+k-1)$  is the Pochhammer symbol \cite{handbook}, $r$ and $\tilde{r}$ are the non-central parameters, $I_n(\cdot)$ is the $n$-th order modified Bessel function of the first kind, $\Gamma\left(\cdot\right)$ is the Gamma function, and $Q_N\left(\cdot\right)$ is the $N$-th order Marcum Q-function. However, \eqref{ExactWDTP} is too difficult to handle. In order to gain insights, the approximated closed-form of WDT outage probability is presented as follows.
	\begin{theorem}\label{th1}
		The WDT outage probability with the WDT oriented strategy for maximizing SINR is approximated as 
		\begin{equation}\label{approximateWDTi}
			\epsilon_{\text{WDT-SINR}}^i\approx	\left[1-K\left(\frac{\mu^2}{\gamma_{\text{th}}+1}\right)^{N-1}-K\mathcal{C}\right]^{+},
		\end{equation}
		where
		\begin{equation}\label{WDTC}
			\begin{aligned}
				\mathcal{C}=&\left(\frac{2\gamma_{\text{th}}\left(1-\mu^2\right)+1}{2\gamma_{\text{th}}^2+\left(3-\mu^2\right)\gamma_{\text{th}}+1}\right)^{N-1}\left(1-\mu^2\right)\sum_{k=0}^{N-2}\sum_{j=0}^{N-k-2}\gamma_{\text{th}}^{j}\\
				&\enspace(\gamma_{\text{th}}+1)^{k+1}\frac{\left(N-j-k-1\right)_j}{j!}\frac{\mu^{2\left(j+k\right)}}{\left[\left(1-\mu^2\right)\gamma_{\text{th}}+1\right]^{j+k+1}},
			\end{aligned}
		\end{equation}
		and $(a)^{+}=\max(0,a)$.
	\end{theorem}
	\begin{IEEEproof}
		See Appendix \ref{proofth1}.
	\end{IEEEproof}

		Theorem 1 has significantly reduced the computational complexity compared to \eqref{ExactWDTP} by eliminating integrals. Besides, a new  corollary is provided to further simplify Theorem \ref{th1}, as shown below.
		\begin{corollary}\label{WDTapproximatefurther}
			The WDT outage probability with the WDT oriented strategy can be further simplified as
			\begin{equation}\label{FurtherWDT}
				\epsilon_{\text{WDT-SINR}}^i\approx
				\left[1-K\left(\frac{\mu^2}{\gamma_{\text{th}}+1}\right)^{N-1}-K\left(\frac{1-\mu^2}{\gamma_{\text{th}}+1}\right)^{N-1}\right]^{+}.
			\end{equation}
		\end{corollary}
		\begin{IEEEproof}
			When $\gamma$ is large and $\mu$ is small, the $\mathcal{C}$ in Theorem 1 can be further simplified as 
			\begin{equation}\label{mathcalC}
				\begin{aligned}
					\mathcal{C}&
					\overset{(a)}{\approx}\left(\frac{2\gamma_{\text{th}}\left(1-\mu^2\right)+1}{2\gamma_{\text{th}}^2+\left(3-\mu^2\right)\gamma_{\text{th}}+1}\right)^{N-1}\\
					&\qquad\times\sum_{k=0}^{N-2}\sum_{j=0}^{N-k-2}\frac{\left(N-j-k-1\right)_j}{j!}\left(\frac{\mu^{2}}{1-\mu^2}\right)^{j+k}\\
					&\overset{(b)}{\approx}\left(\frac{1-\mu^2}{\gamma_{\text{th}}+1}\right)^{N-1}
				\end{aligned}
			\end{equation}
			where $(a)$ accounts for the approximation that $1+\left(1-\mu^2\right)\gamma_{\text{th}}\approx \left(1-\mu^2\right)\gamma_{\text{th}}$, $1+\gamma_{\text{th}} \approx \gamma_{\text{th}}$, and $(b)$ neglects the higher-order terms which get	smaller and applies the assumption that $\gamma_{\text{th}}$ is large and $\mu$ is small. Then, the proof of Corollary \ref{WDTapproximatefurther} ends.
		\end{IEEEproof}
		\begin{remark}
			It can be observed from \eqref{FurtherWDT} that the WDT outage probability with WDT oriented port selection strategy increases when we increase  $N$, $\gamma_{\text{th}}$ or decrease $K$.
	\end{remark}

	The results in \eqref{ExactWDTP}, \eqref{approximateWDTi} and \eqref{FurtherWDT} provide exact and approximated expressions for the WDT outage probability with the WDT oriented strategy of the FAMA-assisted IDET system. In what follows, we investigate the WET outage, which is defined as the event that the EHP at UE $i$ falls below a certain threshold $Q_{\text{th}}$. To start with, we define
	\begin{equation}\label{defineQki}
		\begin{aligned}
			\widehat{Q}_{k}^i&=\sum_{m=1 }^{N} \left[\left(x_k^{(m,i)}+\frac{\mu x_0^{(m,i)}}{\sqrt{1-\mu^2}} \right)^2+\right.\\
			&\qquad\qquad\qquad\left.\left(y_k^{(m,i)}+\frac{\mu y_0^{(m,i)}}{\sqrt{1-\mu^2}} \right)^2\right]\\
			&\triangleq X_k+Y_k.
		\end{aligned}
	\end{equation}
	Using the port selection strategy in \eqref{SINRM}, the WET outage probability with the WDT oriented strategy is then given by
		\begin{equation}\label{orignialMSC}
			\begin{aligned}
				\epsilon^i_{\text{WET-SINR}}&=\text{Prob}\left(Q_{k^{\ast}}^{i}<Q_{\text{th}}\vert k^{\ast}=\arg\max_k \text{SINR}_k^i=\frac{X_k}{Y_k} \right)\\
				&=\text{Prob}\left(X_{k^{\ast}}+Y_{k^{\ast}}<\widehat{Q}_{\text{th}}\vert k^{\ast}=\arg\max_k  \frac{X_k}{Y_k} \right)\\
				&=\frac{\text{Prob}\left(X_{k^{\ast}}+Y_{k^{\ast}}<\widehat{Q}_{\text{th}}, k^{\ast}=\arg\max_k \frac{X_k}{Y_k}\right)}{\text{Prob}\left(k^{\ast}=\arg\max_k \frac{X_k}{Y_k}\right)}\\
				&\overset{(a)}{=}K\text{Prob}\left(X_{k^{\ast}}+Y_{k^{\ast}}<\widehat{Q}_{\text{th}}, k^{\ast}=\max_k \frac{X_k}{Y_k}\right)\\
				&=K\text{Prob}\left( \max_{n\neq {k^{\ast}}} \frac{X_{n}}{Y_{n}}<\frac{X_{k^{\ast}}}{Y_{k^{\ast}}} <\frac{\widehat{Q}_{\text{th}}}{Y_{k^{\ast}}}-1\right)\\
				&=K\text{Prob}\left(Y_{k^{\ast}} \max_{n\neq {k^{\ast}}} \frac{X_n}{Y_n}<X_{k^{\ast}}<\widehat{Q}_{\text{th}}-Y_{k^{\ast}}\right),
			\end{aligned}
		\end{equation}
		where $(a)$ holds because the correlation coefficient between different ports is a constant and thus the channel gain of all the ports follows the same distribution, $\widehat{Q}_{\text{th}}=\frac{d^\beta Q_{\text{th}}}{(1-\mu^2)(1-\rho_i)P}$.
	
	Next, we present the exact WET outage probability with the WDT oriented strategy in Theorem \ref{th2} below.
	
	\begin{theorem}\label{th2}
		The WET outage probability with the WDT oriented port selection strategy is expressed as \eqref{WETMSC} (see top of next page), 
		\begin{figure*}[]
			% Store the current equation number.
			%		\setcounter{MYtempeqncnt}{\value{equation}}
			%		\setcounter{equation}{18}
			\begin{equation}\label{WETMSC}
				\begin{aligned}
					\epsilon^i_{\text{WET-SINR}}= K\int_{r_1=0}^{\infty}\int_{r_2=0}^{\infty}\int_{z=0}^{\infty}\int_{y=0}^{\frac{\widehat{Q}_{\text{th}}}{1+z}}\frac{r_2^{N-2}\exp{(-\frac{r_1+r_2}{2}})}{2^{N+1}\Gamma(N-1)} \left[Q_1\left(\sqrt{\frac{\mu^2}{1-\mu^2}r_1},\sqrt{yz}\right)-Q_1\left(\sqrt{\frac{\mu^2}{1-\mu^2}r_1},\widehat{Q}_{\text{th}}\right)\right]\times\\
					\left(\frac{\left(1-\mu^2 \right) y}{\mu^2r_2}\right)^{\frac{N-2}{2}} \exp\left( -\frac{y+\frac{\mu^2}{1-\mu^2}r_2}{2}\right) I_{N-2}\left(\sqrt{\frac{\mu^2r_2y}{1-\mu^2}}\right)\mathcal{D}(r_1,r_2,z)dydzdr_1dr_2\\				
				\end{aligned}
			\end{equation}
			\hrulefill
			% Restore the current equation number.
			%\setcounter{equation}{\value{MYtempeqncnt}}
			% The IEEE uses as a separator
			% The spacer can be tweaked to stop underfull vboxes.
		\end{figure*}
		where $Q_1\left(\cdot\right)$ is the marcum-Q function of order 1.
		\begin{equation}
			\begin{aligned}
				\mathcal{D}&(r_1,r_2,z)=(K-1)\left[1-\frac{1}{2}\int_{0}^{\infty}Q_1\left(\sqrt{\frac{\mu^2r_1}{1-\mu^2}},\sqrt{zy}\right)\right.\\
				&\left.\times  e^{-\frac{y+\frac{\mu^2}{1-\mu^2}r_2}{2}}I_{N-2}\left(\sqrt{\frac{\mu^2r_2y}{1-\mu^2}}\right)dy\right]^{K-2} \times\\
				&\int_{0}^{\infty}\frac{y^{\frac{N}{2}}}{4\left(\frac{\mu^2}{1-\mu^2}r_2\right)^{\frac{N-2}{2}}}e^{-\frac{(z+1)y+\frac{\mu^2}{1-\mu^2}(r_1+r_2)}{2}}\times\\
				&\qquad\qquad I_0\left(\sqrt{\frac{\mu^2r_1zy}{1-\mu^2}}\right) I_{N-2}\left(\sqrt{\frac{\mu^2r_2y}{1-\mu^2}}\right)dy.				
			\end{aligned}
		\end{equation}
	\end{theorem}
	
	\begin{IEEEproof}
		See Appendix \ref{proofth2}.
	\end{IEEEproof}
	It can be seen that the expression in \eqref{WETMSC} is too complex to gain insights. Thus, we introduce the following lemma to approximate the outage probability. 
		\begin{lemma}\label{lemma2}
			When $\mu=0$, the variables $\frac{X_k}{Y_k}$ and $X_k+Y_k$ are  independent.
		\end{lemma}
		\begin{IEEEproof}
			See Appendix \ref{prooflemma1}.
	\end{IEEEproof}
	
	Using the  Lemma \ref{lemma2}, the WDT outage probability under the WET oriented strategy is approximated in the following corollary.
		\begin{corollary}\label{corollary:approximateWET-SINR}
			The WET outage probability under the WDT oriented port selection strategy is approximated as
			\begin{equation}\label{approximateWET-SINR}
				\epsilon^i_{\text{WET-SINR}}\approx \frac{1}{\Gamma\left(N\right)}\gamma\left(N,\widetilde{Q}_{\text{th}}/2\right),
			\end{equation}
			where $\gamma(\cdot,\cdot)$ is the lower incomplete gamma function and $\widetilde{Q}_{\text{th}}=\lim\limits_{\mu\rightarrow0}\widehat{Q}_{\text{th}}=\frac{d^\beta Q_{\text{th}}}{(1-\rho_i)P}$.
		\end{corollary}
		\begin{IEEEproof}
			It was demonstrated that the correlation coefficient $\mu$ decreases with the increase of $W$ \cite{sfama}. Thus, $X_k$ and $Y_k$ can be approximated as \eqref{approximateX_k} and \eqref{approximateY_k}, respectively. Then, the WET outage probability with the WDT oriented port selection strategy can be approximated as
			\begin{equation}
				\begin{aligned}
					\epsilon^i_{\text{WET-SINR}}&=\text{Prob}\left(Q_{k^{\ast}}^{i}<Q_{\text{th}}\left| k^{\ast}=\arg\max_k \text{SINR}_k^i=\frac{X_k}{Y_k} \right.\right)\\
					&=\text{Prob}\left(X_{k^{\ast}}+Y_{k^{\ast}}<\widehat{Q}_{\text{th}}\left| k^{\ast}=\arg\max_k  \frac{X_k}{Y_k} \right.\right)\\
					&\overset{(a)}{=}\text{Prob}\left(X_{k^{\ast}}+Y_{k^{\ast}}<\widehat{Q}_{\text{th}}\right),\forall k^{\ast}.
				\end{aligned}
			\end{equation}
			where $(a)$ accounts for the result of Lemma \ref{lemma2} that the $X_k+Y_k$ and $X_k/Y_k$ can be regarded as independent when $\mu$ is small.
			Note that $Z_{k^{\ast}}=\widehat{X}_{k^{\ast}}+\widehat{Y}_{k^{\ast}}$ follows a central Chi-square distribution with CDF
			\begin{equation}
				F_{Z_{k^{\ast}}}\left(z\right)=\frac{1}{\Gamma\left(N\right)}\gamma\left(N,z/2\right).
			\end{equation}
			Thus, the \eqref{approximateWET-SINR} can be readily obtained, completing the proof.
		\end{IEEEproof}
		\begin{remark}
			Corollary \ref{corollary:approximateWET-SINR} provides a closed-form expression for the WET outage probability under the WDT oriented port selection strategy, showing that the results are mainly dependent with $N$ and $\widetilde{Q}_{\text{th}}$.
		\end{remark}
	
	\subsection{Outage probabilities analysis with WET oriented port selection strategy}
	In what follows, we investigate the outage probabilities of
	WET and WDT with the WET oriented strategy for maximizing
	the EHP in \eqref{HPM}. Firstly, the WET outage probability is then expressed as
	\begin{align}
		\epsilon^i_{\text{WET-EHP}}&=\text{Prob}\left( \max_k \enspace Q_k^i < Q_{\text{th}}\right)\notag\\
		&=\text{Prob}\left(\max_k\enspace \widehat{Q}_{k}^i=X_k+Y_k < \widehat{Q}_{\text{th}}\right).\label{WEToutageexpression}
	\end{align}
	According to \cite{lin}, the WET outage probability with the WET oriented strategy is expressed by	
	\begin{multline}\label{exactWETP}
		\epsilon^i_{\text{WET-EHP}}=\int_{0}^{\infty}\frac{r^{N-1}e^{-\frac{r}{2}}}{2^N\Gamma(N)}\\\times
		\left[1-Q_N(\sqrt{\frac{\mu^2r}{1-\mu^2}},\sqrt{\widehat{Q}_{\text{th}}})\right]^Kdr.
	\end{multline}
	
	Next, \eqref{exactWETP} is approximated in the following theorem.
	\begin{theorem}\label{th3}
		The WET outage probability with the WET oriented strategy for maximizing the EHP is approximated as
		\begin{multline}\label{ApproximateWET}
			\epsilon^i_{\text{WET-EHP}}\approx\left[1-K  \frac{\Gamma\left(N,\frac{\widehat{Q}_{\text{th}}}{2}\right)}{\Gamma(N)}-K\frac{\mu^2\left(\frac{\widehat{Q}_{\text{th}}}{2}\right)^{N}e^{-\frac{\widehat{Q}_{\text{th}}}{2}}}{N!}\right.\\
			\left.\times\sum_{l=0}^{N-1}(1-\mu^2)^l{_1F_1}(l+1;N+1;\frac{\mu^2\widehat{Q}_{\text{th}}}{2})\right]^{+},
		\end{multline}
		where $\Gamma(\cdot,\cdot)$ is the upper incomplete gamma function.
	\end{theorem}
	
	\begin{IEEEproof}
		See Appendix \ref{proofth3}.
	\end{IEEEproof}
		\begin{remark}
			Note that in \eqref{exactWETP},  when $W$ is large ($\mu$ is small), we have $Q_N(\sqrt{\frac{\mu^2}{1-\mu^2}},\sqrt{\widehat{Q}_{\text{th}}})\approx \frac{\Gamma\left(N,\widehat{Q}_{\text{th}}/2\right)}{\Gamma\left(N\right)}$, which implies that the WET outage probability with WET oriented port selection strategy saturates and does not continue to decrease. In this case, the WET outage probability with WET oriented port selection strategy in \eqref{exactWETP} is determined by $K$, $Q_{\text{th}}$, and $N$. Besides, Theorem \ref{th3}  has reduced the computational complexity compared to \eqref{exactWETP} by eliminating integrals. 
	\end{remark}
	In what follows, we investigate the WDT outage probability with the WET oriented port selection strategy, which is given by
	\begin{equation}
		\begin{aligned}
			&\epsilon_{\text{WDT-EHP}}^i\\
			&=\text{Prob}\left(\text{SINR}_{k^{\ast}}^i=\frac{X_{k^{\ast}}}{Y_{k^{\ast}}}<\gamma_{\text{th}}\vert k^{\ast}=\arg\max_k \widehat{Q}_{k}^i \right)\\
			&=\frac{\text{Prob}\left(\frac{X_{k^{\ast}}}{Y_{k^{\ast}}}<\gamma_{\text{th}}, k^{\ast}=\arg\max_k X_k+Y_k\right)}{\text{Prob}\left(k^{\ast}=\arg\max_k X_k+Y_k\right)}\\
			&=K\text{Prob}\left(\frac{X_{k^{\ast}}}{Y_{k^{\ast}}}<\gamma_{\text{th}}, k^{\ast}=\arg\max_k X_k+Y_k\right)\\
			&=K\text{Prob}\left(\frac{X_{k^{\ast}}}{Y_{k^{\ast}}}<\gamma_{\text{th}},X_{k^{\ast}}+Y_{k^{\ast}}\geq\max_{n\neq {k^{\ast}}} X_n+Y_n\right)\\
			&=K\text{Prob}\left(\max_{n\neq {k^{\ast}}} X_n+Y_n-Y_{k^{\ast}} <X_{k^{\ast}}<\gamma_{\text{th}} Y_{k^{\ast}}\right).
		\end{aligned}
	\end{equation}
	
	\begin{theorem}\label{th4}
		The WDT outage probability with the WET oriented port strategy is given by \eqref{WDTMHP} (see top of next page).
		\begin{figure*}[]
			\begin{equation}\label{WDTMHP}
				\begin{aligned}
					\epsilon^i_{\text{WDT-EHP}}&=K\int_{r_1=0}^{\infty}\int_{r_2=0}^{\infty}\int_{y=0}^{\infty}\int_{t=0}^{\gamma_{\text{th}} y}\frac{r_2^{N-2}e^{-\frac{r_1+r_2}{2}}}{2^{N+2}\Gamma(N-1)}\left[1-Q_N\left(\sqrt{\frac{\mu^2}{1-\mu^2}(r_1+r_2)},\sqrt{t+y}\right)\right]^{K-1}\\
					&\qquad\times e^{-\frac{\frac{\mu^2}{1-\mu^2}r_1+t}{2}-\frac{y+\frac{\mu^2}{1-\mu^2}r_2}{2}} I_0\left(\sqrt{\frac{\mu^2r_1t}{1-\mu^2}}\right)\times\left(\frac{\left(1-\mu^2\right)y}{\mu^2r_2}\right)^{\frac{N-2}{2}}I_{N-2}\left(\sqrt{\frac{\mu^2r_2y}{1-\mu^2}}\right)dtdydr_1dr_2
				\end{aligned}
			\end{equation}
			\hrulefill
			% Restore the current equation number.
			%\setcounter{equation}{\value{MYtempeqncnt}}
			% The IEEE uses as a separator
			% The spacer can be tweaked to stop underfull vboxes.
		\end{figure*}
	\end{theorem}
	
	\begin{IEEEproof}
		See Appendix \ref{proofth4}.
	\end{IEEEproof}
		\begin{corollary}\label{corollary:WDT-EHP}
			The WDT outage probability under the WET oriented strategy with larger $W$ is approximated as
			\begin{equation}\label{approximate:WDTEHP}
				\epsilon_{\text{WDT-EHP}}^i\approx 1-\frac{1}{\left(\gamma_{\text{th}}+1\right)^{N-1}}.
			\end{equation}
		\end{corollary}
		\begin{IEEEproof}
			Similarly, when the antenna size $W$ is large, $X_k$ and $Y_k$ can be approximated as
			\eqref{approximateX_k} and \eqref{approximateY_k}, respectively. Then, the WDT outage probability is approximated as
			\begin{equation}
				\begin{aligned}
					\epsilon_{\text{WDT-EHP}}^i&=\text{Prob}\left(\text{SINR}_{k^{\ast}}^i=\frac{X_{k^{\ast}}}{Y_{k^{\ast}}}<\gamma_{\text{th}}\vert k^{\ast}=\arg\max_k \widehat{Q}_{k}^i \right)\\
					&=\text{Prob}\left(\frac{X_{k^{\ast}}}{Y_{k^{\ast}}}<\gamma_{\text{th}}\vert k^{\ast}=\arg\max_k X_k+Y_k \right)\\
					&\overset{(a)}{\approx}\text{Prob}\left(\frac{X_{k^{\ast}}}{Y_{k^{\ast}}}<\gamma_{\text{th}} \right),\forall k^{\ast}.
				\end{aligned}
			\end{equation}
			where $(a)$ accounts for the result of Lemma \ref{lemma2} that the $X_k+Y_k$ and $X_k/Y_k$ can be regarded as independent when $\mu$ is small.
			Note that the $\widehat{X}_{k^{\ast}}$ is exponentially distributed with CDF 
			\begin{equation}
				F_{\widehat{X}_{k^{\ast}}}\left(x\right)=1-e^{-x/2},
			\end{equation}
			and 	$\widehat{Y}_{k^{\ast}}$ is Chi-squared	distributed with PDF in \eqref{approximateYkmu=0}.
			Thus, the outage probability is then calculated as
			\begin{equation}\label{outcalculate}
				\epsilon_{\text{WDT-EHP}}^i=\int_{0}^{\infty}F_{\widehat{X}_{k^{\ast}}\vert\widehat{Y}_{k^{\ast}}}\left(\gamma_{\text{th}} y\right)f_{\widehat{Y}_{k^{\ast}}}\left(y\right)dy.
			\end{equation}
			Finally, \eqref{approximate:WDTEHP} can be obtained by following the same steps as in \cite[eq. 63]{sfama}, completing 
			the proof of Corollary \ref{corollary:WDT-EHP}.
		\end{IEEEproof}
		\begin{remark}
			Corollary \ref{corollary:WDT-EHP} provides a closed-form expression for the WET outage probability under the WDT oriented strategy, showing that the results are mainly dependent with $N$ and $\gamma_{\text{th}}$. 
	\end{remark}
	\subsection{Analysis of multiplexing gain for WDT and WET}
	It is assumed that each BS antenna transmits at a constant rate, and ID is successfully performed with the probability of ($1-\epsilon_{\text{WDT-SINR}}^i$). It is also assumed that  all the UEs are statistically identical and have the same SINR threshold  or EHP threshold. It is also assumed that  all the UEs are statistically identical and have the same SINR threshold. According to \cite{sfama}, the WDT multiplexing gain of the FAMA-assisted IDET system is defined as
	\begin{equation}\label{mWDT}
		m_{\text{WDT}}=N\left(1-\epsilon_{\text{WDT-SINR}}^i\right).
	\end{equation}

	Similarly, we can define the WET multiplexing gain as
	\begin{equation}\label{mWET}
		m_{\text{WET}}=N\left(1-\epsilon_{\text{WET-EHP}}^i\right),
	\end{equation}
	which adopts similar assumptions as in \eqref{mWDT}.
	\subsection{Analysis of energy efficiency}
		The sum rate of the system is 
		\begin{equation}
			R=\sum_{i=1}^{N}B\log_2\left( 1+\text{SINR}_{k_i}^i\right),
		\end{equation}
		where $B$ is the channel bandwidth of the system, and $\text{SINR}_{k_i}^i$  denotes that the received SINR at the $k_i$-th port of UE $i$, which is given as \eqref{SINR}. 
		
		And the harvested energy of the system is denoted as
		\begin{equation}
			Q_{\text{harvest}}=\sum_{i=1}^N Q_{k_i}^i,
		\end{equation}
		where $Q_{k_i}^i$ denotes that the harvested power at the $k_i$-th port of UE $i$, which is given as \eqref{Qk}.
		Since it is assumed that the transmit power of all the BS antennas is the same, the total power consumption of the system is formulated as
		\begin{equation}
			Q_{\text{total}}=NP+P_C-\sum_{i=1}^N Q_{k_i}^i,
		\end{equation}
		where $P_C$ denotes the fixed power consumption of the
		system hardware.
		\begin{definition}
			Energy Efficiency (EE) of the FAMA-assisted IDET system is defined as the ratio
			of the sum rate to the total power consumption which is
			expressed by
		\end{definition}
		\begin{equation}
			\text{EE}\triangleq \frac{R}{Q_{\text{total}}}=\frac{\sum_{i=1}^{N}B\log_2\left( 1+\text{SINR}_{k_i}^i\right)}{NP+P_C-\sum_{i=1}^N Q_{k_i}^i},
		\end{equation}
		where $k_i$ can be determined by the WDT oriented port selection strategy in \eqref{SINRM} or the WET oriented port selection strategy in \eqref{HPM}.
	\section{IDET Performance Analysis}\label{section4}
	We have provided outage probabilities with WDT oriented port selection strategy in III-B and  with  WET oriented port selection strategy in III-C, respectively. However, in order to measure whether the system functions well, the analysis of IDET performance is very necessary because it provides a more complete picture of the system performance compared to considering specific port selection strategies. Two kinds of IDET outage probabilities are defined as follows. 
	
	\textbf{Special IDET outage probability:} If both WDT and WET suffer from an outage, then IDET suffers from a special outage, indicating that the received SINR and EHP at all the ports are both lower than their respective thresholds. This corresponds to the worst case of the IDET system. Then, the special IDET outage probability is formulated as
	\begin{equation}\label{IDETdefine1}
		\begin{aligned}
			&\epsilon_{\text{{IDET-special}}}^i\\
			&=\text{Prob}\left(\text{SINR}_k^i < \gamma_{\text{th}}, Q_k^i < Q_{\text{th}}, \forall k= 1, \dots, K\right)\\
			&=\text{Prob}\left(\frac{X_k}{Y_k}<\gamma_{\text{th}}, X_k+Y_k<\widehat{Q}_{\text{th}}, \forall k= 1, \dots, K\right)\\
			&=\text{Prob}\left(\frac{X_k}{\gamma_{\text{th}}}< Y_k<\widehat{Q}_{\text{th}}-X_k, \forall k= 1, \dots, K\right).
		\end{aligned}
	\end{equation}
	
	Next, we present the expression for the special IDET outage probability in Theorem \ref{th5} as follows.
	
	\begin{theorem}\label{th5}
		The special IDET outage probability is formulated in \eqref{finalIDET} (see top of next page).
		\begin{figure*}[]
			% Store the current equation number.
			%			\setcounter{MYtempeqncnt}{\value{equation}}
			% Set the equation number to one less than the one
			% desired for the first equation here.
			% The value here will have to changed if equations
			% are added or removed prior to the place these
			% equations are referenced in the main text.
			%			\setcounter{equation}{28}
			\begin{equation}\label{finalIDET}
				\begin{aligned}
					&\epsilon_{\text{IDET-special}}^i=\int_{r_1=0}^{\infty}\int_{r_2=0}^{\infty}\frac{r_2^{N-2}e^{-\left(-\frac{r_1+r_2}{2}\right)}}{2^{N}\Gamma(N-1)} \left\{\int_{x=0}^{\frac{\widehat{Q}_{\text{th}}}{\left(1+1/\gamma\right)}} \left[Q_{N-1}\left(\sqrt{\frac{\mu^2r_1}{1-\mu^2}},\sqrt{\frac{x}{\gamma}}\right)-Q_{N-1}\left(\sqrt{\frac{\mu^2r_1}{1-\mu^2}},\sqrt{\widehat{Q}_{\text{th}}-x}\right)\right]\right.\\ 
					&\quad\quad\quad\quad\quad\quad \left.\times \frac{1}{2}e^{-\frac{x+\frac{\mu^2}{1-\mu^2}r_2}{2}}I_0\left(\sqrt{\frac{\mu^2}{1-\mu^2}r_2x}\right)dx\right\}^Kdr_1dr_2
				\end{aligned}
			\end{equation}
			\hrulefill
			% Restore the current equation number.
			%\setcounter{equation}{\value{MYtempeqncnt}}
			% The IEEE uses as a separator
			% The spacer can be tweaked to stop underfull vboxes.
		\end{figure*}
	\end{theorem}
	
	\begin{IEEEproof}
		See Appendix F.
	\end{IEEEproof}
	
	We have provided an accurate expression for evaluating the IDET outage probability in Theorem \ref{th5}. Nonetheless, the expression is hard to handle. In order to approximate it into a closed form, we first introduce Lemma \ref{lemma2} below.
	
	Afterwards, by applying Lemma \ref{lemma2}, the special IDET outage probability is formulated as
	\begin{equation}\label{approximateIDET}
		\begin{aligned}
			&\epsilon_{\text{{IDET-special}}}^i\\
			&\hspace{1.5mm}=\text{Prob}\left(\max_k \frac{X_k}{Y_k}<\gamma_{\text{th}}, \max_k  X_k+Y_k<\widehat{Q}_{\text{th}}\right)\\
			&\overset{\mu\rightarrow0}{\approx}\text{Prob}\left(\max_k \frac{X_k}{Y_k}<\gamma_{\text{th}}\right)\text{Prob}\left(\max_k X_k+Y_k<\widehat{Q}_{\text{th}}\right)\\
			&\hspace{1.5mm}=\epsilon_{\text{{WDT-SINR}}}^i\epsilon_{\text{{WET-EHP}}}^i,
		\end{aligned}
	\end{equation}
	where
	$\epsilon_{\text{{WDT-SINR}}}^i$ is accurately derived in \eqref{ExactWDTP} and approximated in Theorem \ref{th1}, while $\epsilon_{\text{{WET-EHP}}}^i$ is accurately derived in \eqref{exactWETP} and approximated in Theorem \ref{th3}. Thus, $\epsilon_{\text{{IDET-Special}}}^i$ and $\epsilon_{\text{{IDET-General}}}^i$ can also be approximated in closed forms.
	
	\begin{corollary}\label{corollary3}
		For small $N$, the special IDET outage probability can be further simplified as
		\begin{equation}\label{insightIDET1}
			\epsilon_{\text{IDET-special}}^i\approx\epsilon_{\text{WDT-SINR}}^i,
		\end{equation}
		while for large $N$, the special IDET outage probability can be further simplified as
		\begin{equation}\label{insightIDET2}
			\epsilon_{\text{IDET-special}}^i\approx\epsilon_{\text{WET-EHP}}^i.
		\end{equation}
	\end{corollary}
	
	\begin{IEEEproof}
		See Appendix G.
	\end{IEEEproof}
	\begin{remark}
		Corollary \ref{corollary3} illustrates that when the number of BS antenna-UE pairs is relatively small, the special IDET outage probability is mainly determined by WDT. On the other hand, when the number of BS antenna-UE pairs is relatively large, the special IDET outage probability is primarily determined by WET. In other words, there is a trade-off between WET and WDT with respect to the number of BS antenna-UE pairs.
		
	\end{remark}
	
	\textbf{General IDET outage probability:} If either WDT or WET suffers from an outage, then the system suffers from a general IDET outage, indicating that no ports can be selected to satisfy both the WDT and WET requirements. It can be used to further assess the performance of successful transmission of both WDT and WET, which is described by
	\begin{equation}\label{IDETdefine2}
		\begin{aligned}
			&\epsilon_{\text{{IDET-general}}}^i\\
			&\hspace{.5mm}=\text{Prob}\left( \left\{ \max_k \text{SINR}_k^i < \gamma_{\text{th}}\right\} \cup \left\{\max_k Q_k^i < Q_{\text{th}}\right\}\right)\\
			&\overset{(a)}{=}\epsilon_{\text{{WDT-SINR}}}^i+\epsilon_{\text{{WET-EHP}}}^i-\epsilon_{\text{{IDET-special}}}^i.
		\end{aligned}
	\end{equation} 
	where $(a)$ uses the addition law of probability. Similarly, since $\epsilon_{\text{{WDT-SINR}}}^i$ is derived in \eqref{ExactWDTP} and approximated in Theorem \ref{th1}, $\epsilon_{\text{{WET-EHP}}}^i$ is derived in \eqref{exactWETP} and approximated in Theorem \ref{th3}, and $\epsilon_{\text{{IDET-special}}}^i$ is derived in Theorem \ref{th5} and approximated in \eqref{approximateIDET}, $\epsilon_{\text{{IDET-general}}}^i$ can be approximated in the closed form.
	
	Similarly, the IDET multiplexing gain for the special case and the general case are analyzed as 
	\begin{align}
		m_{\text{IDET-special}}&=N(1-\epsilon_{\text{IDET-special}}^i),\label{specialIDETmultiplexing}\\
		m_{\text{IDET-general}}&=N(1-\epsilon_{\text{IDET-general}}^i),\label{generalIDETmultiplexing}
	\end{align}
	where we have assumed that each BS antenna transmits at a constant rate. It is also assumed that all UEs are statistically identical, sharing the same SINR and EHP thresholds.
		\section{Extended to Rician Channels}\label{section7}
		In this section, a more general Rician channel that includes the LoS component is conceived to analyze the performance of the proposed FAMA-assisted IDET system. A far-field model  is assumed, where all ports in the FAS exhibit the same LoS component. For simplicity, it is assumed that the LoS component is a constant, which is given by \cite{MassiveLuo}
		\begin{equation}\label{ChannelLoS}
			h_{k}^{\left(m,i\right)} =\sqrt{\frac {\kappa\Omega_{m,i}}{\kappa+1}}\bar{h}_k^{\left(m,i\right)}+\sqrt{\frac{\Omega_{m,i}}{2\left( \kappa+1\right) }}\tilde{h}_k^{\left(m,i\right)},
		\end{equation}
		where $\Omega_{m,i}$ is the path loss between the $m$-th BS antenna and UE $i$ with $\Omega_{m,i}=d_{m,i}^{-\beta}$, $\kappa$ denotes the Rician factor, $\bar{h}_k^{\left(m,i\right)}=e^{j\omega^{\left(m,i\right)}}$ is the LoS  component and $\omega^{\left(m,i\right)}$ is the corresponding phase, $\widetilde{h}_k^{\left(m,i\right)}$ is the NLoS component, which is a complex Gaussian random variable and is  given by \eqref{eqgk}. It is noteworthy that \eqref{ChannelLoS} converges to our NLoS channel model in \eqref{eqgk} when $\kappa=0$.  Then, the WDT oriented port selection strategy is given by
		\begin{equation}\label{kSINRrician}
			k_{\text{SINR} }^{\ast}= \arg\max_k \frac{\left\vert h_k^{(i,i)} \right\vert^2 }{\sum_{m\neq i}^{N}\left\vert h_k^{(m,i)} \right\vert^2 }\triangleq \arg\max_k \frac{\widetilde{X}_k}{\widetilde{Y}_k},
		\end{equation}
		and the WET oriented port selection strategy is given by
		\begin{equation}\label{kEHPrician}
			k_{\text{EHP}}^{\ast}= \arg\max_k \enspace \sum_{m=1}^{N}\left\vert h_k^{(m,i)} \right\vert^2\triangleq\arg\max_k \enspace \left(\widetilde{X}_k+\widetilde{Y}_k\right).
		\end{equation}
		In \eqref{kSINRrician} and \eqref{kEHPrician}, we have
		\begin{equation}\label{newX_k}
			\widetilde{X}_k
			=\left(x_k^{\left(i,i\right)}+\frac{\mu\tilde{x}_0^{\left(i,i\right)} }{\sqrt{1-\mu^2}}\right)^2+\left(y_k^{\left(i,i\right)}+\frac{\mu \tilde{y}_0^{\left(i,i\right)}}{\sqrt{1-\mu^2}}\right)^2,\\
		\end{equation}
		\begin{equation}\label{newY_k}
			\widetilde{Y}_k
			=\sum_{m=1\atop m\neq i }^{N}\left(x_k^{\left(m,i\right)}+\frac{\mu\tilde{x}_0^{\left(m,i\right)} }{\sqrt{1-\mu^2}}\right)^2+\left(y_k^{\left(m,i\right)}+\frac{\mu \tilde{y}_0^{\left(m,i\right)}}{\sqrt{1-\mu^2}}\right)^2,
		\end{equation}
		where $\tilde{x}_0^{\left(m,i\right)}\sim\mathcal{CN}(\frac{\sqrt{\kappa}\cos(\omega^{\left(m,i\right)})}{\mu},1 )$, $\tilde{y}_0^{\left(m,i\right)}\sim\mathcal{CN}(\frac{\sqrt{\kappa}\sin(\omega^{(m,i)})}{\mu},1 ), \forall  m=1,\ldots,N$. After that, by defining  $\widetilde{v}_1=(\tilde{x}_0^{\left(i,i\right)})^2+(\tilde{y}_0^{\left(i,i\right)})^2$, and $\widetilde{v}_2=\sum_{m\neq i}^N(\tilde{x}_0^{\left(m,i\right)})^2+(\tilde{y}_0^{\left(m,i\right)})^2$, which follow non-central chi-square distribution with $2$ degrees of freedom and  $2(N-1)$ degrees of freedom, respectively, the PDF of $\widetilde{v}_1$ and $\widetilde{v}_2$ are given by \cite{ProbabilityM.Simon}
		\begin{equation}\label{pdfnewr1}
			f_{\widetilde{v}_1}(v_1)=\frac{1}{2}\exp\left(-\frac{v_1+\frac{\kappa}{\mu^2}}{2}\right)I_{0}\left(\sqrt{\frac{\kappa}{\mu^2}v_1}\right).
		\end{equation}
		\begin{equation}\label{pdfnewr2}
			\begin{aligned}
				f_{\widetilde{v}_2}(v_2)=&\frac{1}{2}\left(\frac{\mu^2 v_2}{\left(N-1\right)\kappa}\right)^{\frac{N-2}{2}}\exp\left(-\frac{v_2+\frac{\left(N-1\right)\kappa}{\mu^2}}{2}\right)\\
				&\times I_{N-2}\left(\sqrt{\frac{\left(N-1\right)\kappa v_2}{\mu^2}}\right).
			\end{aligned}
		\end{equation}
		Then, the WDT outage probability with WDT oriented  port selection strategy is expressed as follows.
		\begin{theorem}
			The exact expression of WDT outage probability with WDT oriented port selection strategy under Rician channel is given by \eqref{ExactWDTPLoS} (see top of this page), where $f_{\widetilde{v}_1}(v_1)$ and $f_{\widetilde{v}_2}(v_2)$ are provided in \eqref{pdfnewr1} and \eqref{pdfnewr2}, respectively.
		\end{theorem}
		\begin{figure*}[]
			\begin{equation}\label{ExactWDTPLoS}
				\begin{aligned}
					\tilde{\epsilon}_{\text{WDT-SINR} }^i&=\int_{0}^{\infty}\int_{0}^{\infty}	f_{\widetilde{v}_1}(v_1)	f_{\widetilde{v}_2}(v_2)\left[ Q_{N-1}\left(\sqrt{\frac{\mu^2\gamma_{\text{th}} v_2}{\left(1-\mu^2\right)\left(\gamma_{\text{th}}+1\right)}},\sqrt{\frac{\mu^2v_1}{\left(1-\mu^2\right)\left(\gamma_{\text{th}}+1\right)}}\right)-e^{-\frac{\mu^2\left(\gamma_{\text{th}} v_2+v_1\right)}{2\left(1-\mu^2\right)\left(\gamma_{\text{th}}+1\right)}}\right.\\
					&\left. \left(\frac{1}{\gamma_{\text{th}}+1}\right)^{N-1}\sum_{k=0}^{N-2}\sum_{j=0}^{N-k-2}\frac{(N-j-k-1)_j}{j!}\left(\frac{v_1}{v_2}\right)^{\frac{j+k}{2}}(\gamma_{\text{th}}+1)^k\gamma_{\text{th}}^{\frac{j-k}{2}}I_{j+k}\left(\frac{\mu^2}{1-\mu^2}\frac{\sqrt{\gamma_{\text{th}} v_1v_2}}{\gamma_{\text{th}}+1}\right)\right]^Kdv_1dv_2
				\end{aligned}
			\end{equation}
			\hrulefill
		\end{figure*}
		\begin{IEEEproof}
			\eqref{ExactWDTPLoS} can be obtained by following the same steps as in \cite[Appendix A]{sfama}, which is omitted here for simplicity.
		\end{IEEEproof}
		
		Then, we define $\widetilde{v}=\widetilde{v}_1+\widetilde{v}_2$, which follows non-central Chi-squared distribution with $2N$ degrees of freedom, while the PDF of $\widetilde{v}$  can be obtained by substituting $\left( N-1\right) $ with $N$ in \eqref{pdfnewr2} as
		\begin{equation}\label{pdfv}
			\begin{aligned}
				f_{\widetilde{v}}(v)=\frac{1}{2}\left(\frac{\mu^2 v}{N\kappa}\right)^{\frac{N-1}{2}}\exp\left(-\frac{v+\frac{N\kappa}{\mu^2}}{2}\right)
				I_{N-1}\left(\sqrt{\frac{N\kappa v}{\mu^2}}\right).
			\end{aligned}
		\end{equation}
		Given the PDF of $\widetilde{v}$, the WET outage probability with WET oriented port selection strategy  is expressed as follows.
		\begin{theorem}
			The exact expression of WET outage probability with WET oriented strategy under Rician channel is given by 
			\begin{equation}\label{exactWETPLoS}
				\begin{aligned}
					\widetilde{\epsilon}^i_{\text{WET-EHP}}=\int_{0}^{\infty}\underbrace{\frac{1}{2}\left(\frac{\mu^2 v}{N\kappa}\right)^{\frac{N-1}{2}}e^{-\frac{r+\frac{N\kappa}{\mu^2}}{2}}I_{N-1}\left(\sqrt{\frac{N\kappa v}{\mu^2}}\right)}_{	f_{\widetilde{v}}(v)}\\
					\times\left[1-Q_N\left(\sqrt{\frac{\mu^2v}{1-\mu^2}},\sqrt{\widehat{Q}_{\text{th}}}\right)\right]^Kdv.\\
				\end{aligned}
			\end{equation}
		\end{theorem}
		\begin{IEEEproof}
			\eqref{exactWETPLoS} can be obtained by following the same steps as in \cite[Theorem 1]{lin}, which is omitted here for simplicity.
		\end{IEEEproof}
		
		Similarly, the WDT outage probability with WET oriented port selection strategy, the WET outage probability with WDT oriented port selection, and the IDET outage probability over Rician channels can be obtained following the same procedures outlined in Appendix B, Appendix E, and Appendix F, respectively, which are all omitted here for simplicity.
	\begin{table}[htbp]
		\centering
		\caption{Parameter Settings}\label{table_parameter}
		\renewcommand\arraystretch{1}
		\begin{tabularx}{\linewidth}{|c|X|c|}
			\hline
			\textbf{Parameter} & \textbf{Description} & \textbf{Values} \\
			\hline \hline
			$d$ & The distance between BS antennas and UEs & 10 m \cite{OntheCoverage} \\
			\hline
			$\beta$ & Path loss exponent & 2 \\
			\hline
			$P$ & Transmit power at BS antennas & 1 W \\
			\hline
			$N$ & Number of UEs & 5 \\
			\hline
			$K$ & Number of ports of a single fluid antenna & 200 \cite{sfama} \\
			\hline
			$W$ & Normalized size of fluid antenna & 5 \cite{sfama} \\
			\hline
			$\rho$ & Power splitting ratio & 0.5 \\
			\hline
			$\gamma_{\text{th}}$ & SINR threshold & 3 dB \cite{sfama} \\
			\hline
			$Q_{\text{th}}$ & EHP threshold & 10 mW \\
			\hline
		\end{tabularx}
	\end{table}
	\section{Numerical Results}\label{section5}
	In this section, the performance of the FAMA-assisted IDET system is evaluated by both the theoretical analysis and Monte-Carlo simulations. It is assumed that the wireless channels of different BS antenna-UE pairs exhibit identical statistical characteristics.  In all our simulations, the markers represent the simulation results, while the curves represent the theoretical results. Unless specifically stated otherwise, the system parameters are set based on the figures in TABLE \ref{table_parameter}. 
	
\begin{figure*}[t]
		\centering
		\begin{minipage}{0.3\textwidth}
			\centering
			\includegraphics[width=\textwidth]{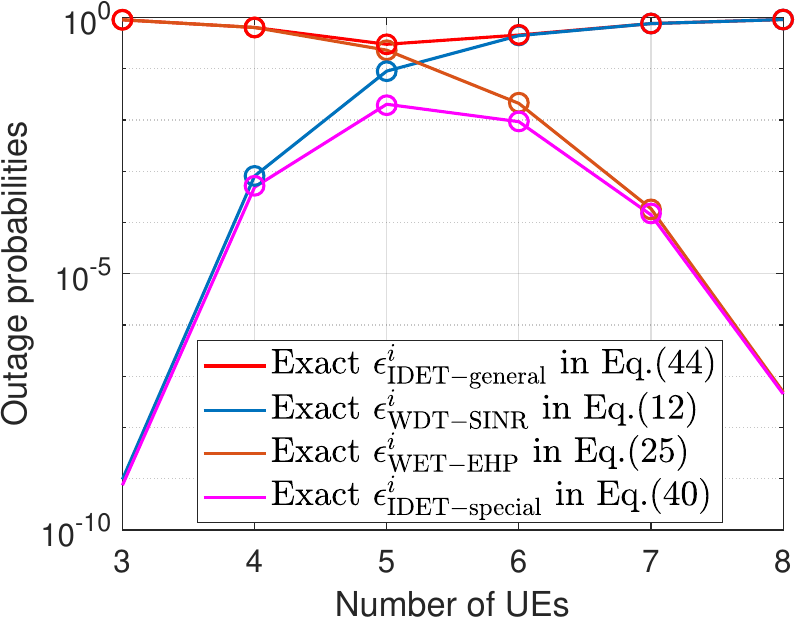}
			\caption{Outage probabilities versus $N$.}
			\label{FigN}
		\end{minipage}\hspace{0.01\linewidth}
		\begin{minipage}{0.3\textwidth}
			\centering
			\includegraphics[width=1\textwidth]{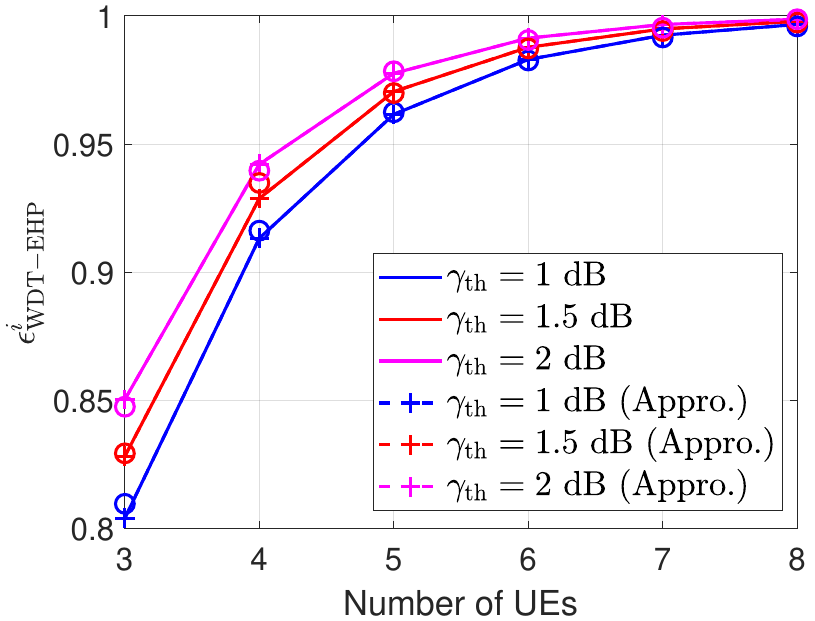}
			\caption{WDT outage probability with the WET oriented port selection strategy versus $N$.}
			\label{WDT-EHPN}
		\end{minipage}\hspace{0.01\linewidth}
		\begin{minipage}{0.3\textwidth}
			\centering
			\includegraphics[width=1\textwidth]{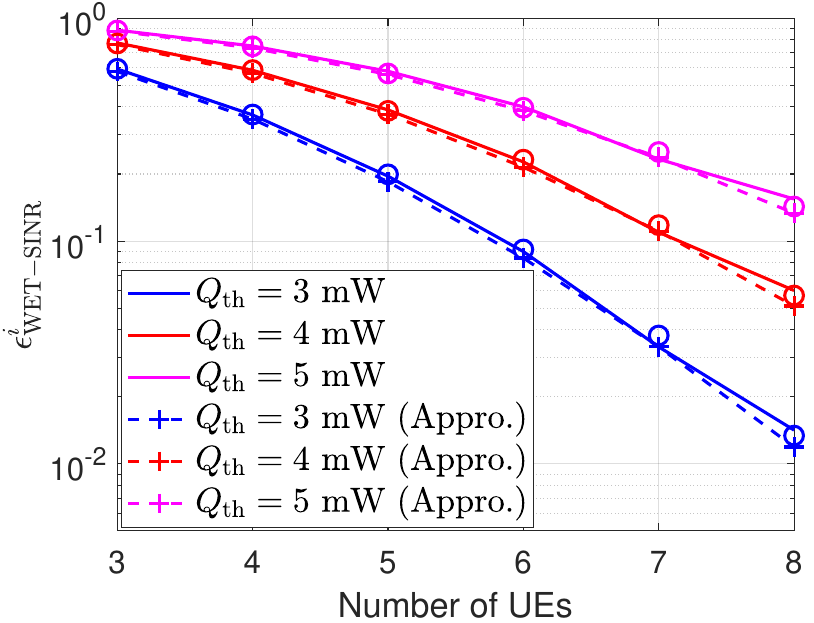}
			\caption{WET outage probability with the WDT oriented port selection strategy versus $N$.}
			\label{WET-SINRN}
		\end{minipage}
		
		\begin{minipage}{0.3\textwidth}
			\centering
			\includegraphics[width=\textwidth]{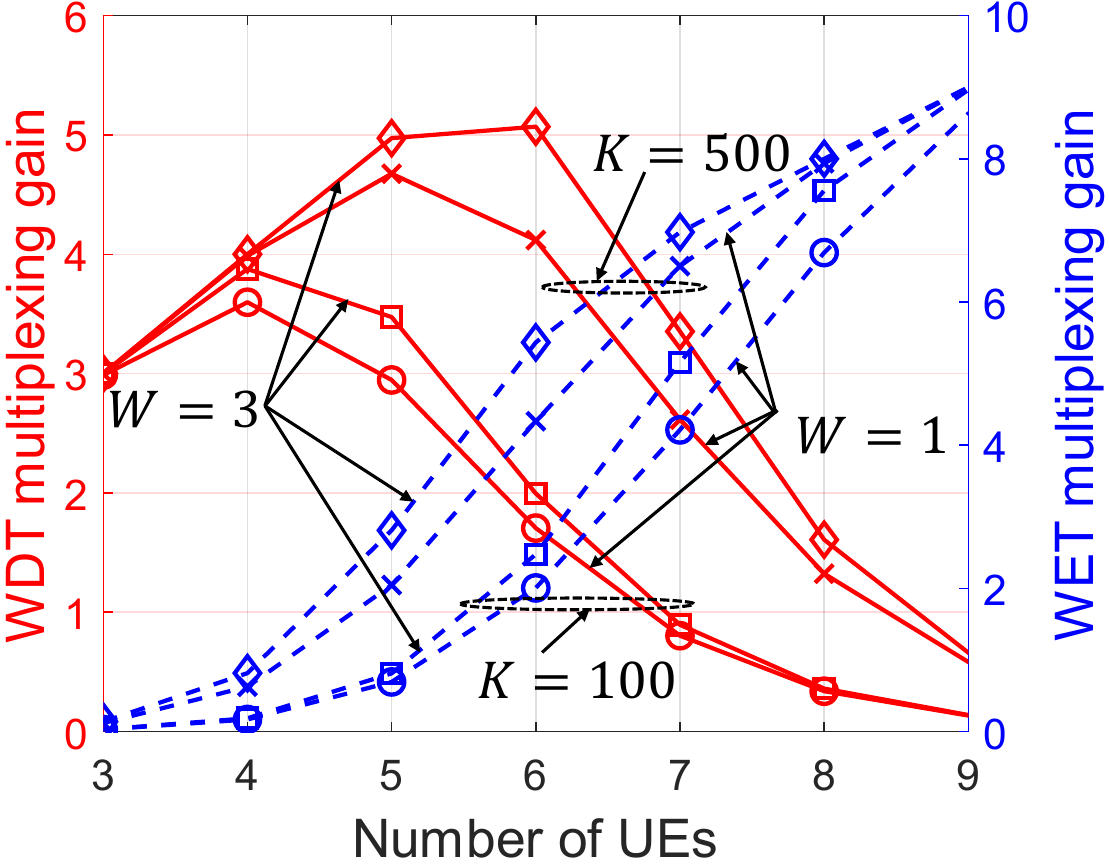}
			\caption{Multiplexing gains of WDT and WET versus $N$; $\gamma_{\text{th}}=3$ dB and $Q_{\text{th}} = 14$ mW.}
			\label{gainWDTWET}
		\end{minipage}\hspace{0.01\linewidth}
		\begin{minipage}{0.3\textwidth}
			\centering
			\includegraphics[width=1\textwidth]{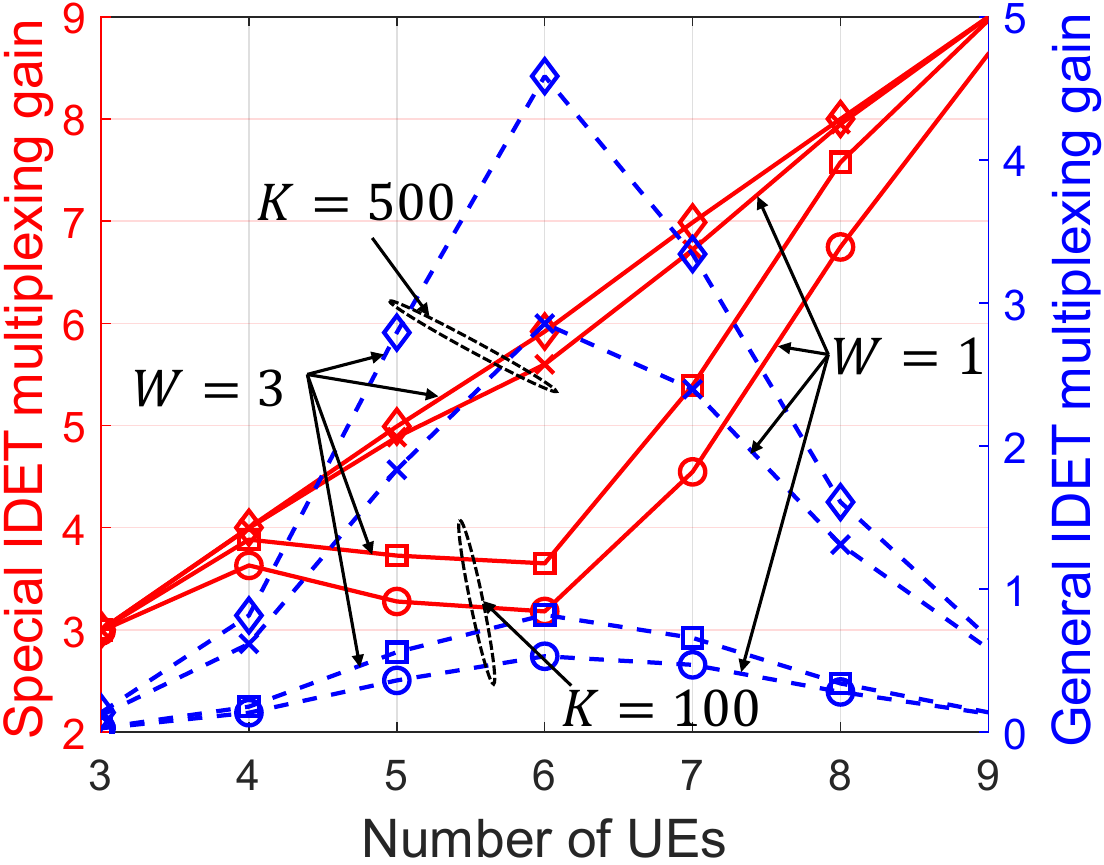}
			\caption{IDET multiplexing gain versus $N$.}
			\label{gainIDET}
		\end{minipage}\hspace{0.01\linewidth}
		\begin{minipage}{0.3\textwidth}
			\centering
			\includegraphics[width=1\textwidth]{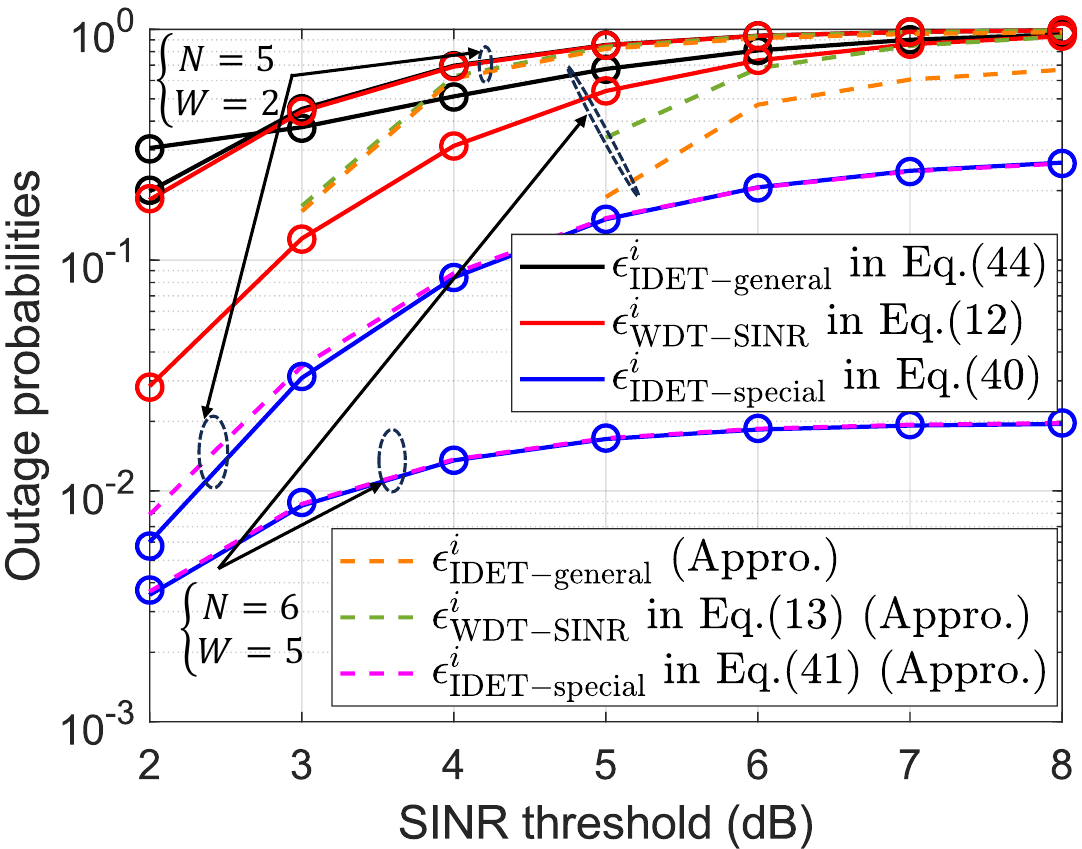}
			\caption{Outage probabilities versus the $Q_{\text{th}}$.}
			\label{FigSINR}
		\end{minipage}
	\end{figure*}

	\cref{FigN,WDT-EHPN,WET-SINRN} illustrate the outage probabilities versus the number $N$ of BS antenna-UE pairs. Specifically, it is seen from \cref{FigN} that the derived analytical results match the simulations very well, which validates our theoretical analysis. As the number of BS antenna-UE pairs $N$ increases, the WET outage probability drops since the UEs can receive more wireless signals from other BS antennas and further glean  more energy. By contrast, the WDT outage probability increases due to more interference from other BS antennas. Besides, we see that the special IDET outage probability converges with the WDT outage probability when $N$ is small, \textit{i.e.}, $N=4$, while converges with the WET outage probability when $N$ is large, \textit{i.e.}, $N=7$. This is in line with Corollary \ref{corollary3}, where $\epsilon_{\text{{IDET-special}}}^i\approx\epsilon_{\text{{WDT-SINR}}}^i$ for small $N$ and $\epsilon_{\text{{IDET-special}}}^i\approx\epsilon_{\text{{WET-EHP}}}^i$ for large $N$. Conversely, the general IDET probability shows a convex trend with respect to $N$. A trade-off between the WDT and WET can be achieved by optimizing the number of UEs.
	
	\cref{WDT-EHPN} and \cref{WET-SINRN} illustrate the WDT outage probability under the WET oriented port selection strategy and WET outage probability under the WDT oriented port selection strategy with respect to $N$, respectively. It is observed that both WDT and WET exhibit poor performance by considering the undesired strategies. This is because when the UEs aim to maximize the SINR via port selection, FAMA tunes into the windows of opportunity where interference naturally disappears in a deep fade, resulting in the lowest EHP. Similarly, when the UEs aim to maximize the EHP via port selection, FAMA tunes into the windows of opportunity with the highest interference, leading to poor performance of WDT. Moreover, both the approximated WET outage probability under the WDT oriented port selection strategy in Corollary \ref{corollary:approximateWET-SINR} and the approximated WDT outage probability under the WET oriented port selection strategy in Corollary \ref{corollary:WDT-EHP} closely match the actual results, demonstrating the validity of our approximation.
	
	The multiplexing gains are presented in \cref{gainWDTWET} and \cref{gainIDET}  to investigate the performance of the FAMA-assisted IDET system versus the number $N$ of BS antenna-UE pairs. Specifically, \cref{gainWDTWET} illustrates that the WDT multiplexing gain is concave with respect to the number $N$ of BS antenna-UE pairs, while the WET multiplexing gain is a strictly increasing function of $N$. This is because when we increase $N$, the WET outage probability reduces, as seen in \cref{FigN}. According to \eqref{mWET}, the WET multiplexing gain increases with $N$. However, as for WDT, a larger $N$ indicates a much higher WDT outage probability, which then reduces the WDT multiplexing gain. Moreover, as shown in \cref{gainIDET}, the special IDET multiplexing gain displays an increase-decrease-increase trend with respect to $N$ when $K=100$. This is in line with Corollary \ref{corollary3}. Therefore, the special IDET multiplexing gain first shows the same concave trend as the WDT multiplexing gain, then shows the same monotonically increasing trend as the WET multiplexing gain. However, for the case with $K=500$, the special IDET multiplexing gain strictly increases with $N$. This is because both the WDT and WET outage probabilities are very small, and therefore the multiplexing gains are mainly determined by $N$. As expected, the general IDET multiplexing gain shows a concave trend with respect to the value of $N$.
	
	Next,  \cref{FigSINR} illustrates the WDT outage probability with the WDT oriented port selection strategy, as well as the IDET outage probabilities without a specific port selection strategy, where different $N$ and $W$ are conceived. As shown in \cref{FigSINR}, the derived approximated closed-form for WDT outage probability closely approximates the theoretical value when the SINR threshold is relatively high, \textit{i.e.}, $\gamma_{\text{th}}=4$ dB. The approximated closed-form expression for general IDET is more accurate at $N=5$ and $W=2$. This is because as $N$ decreases, $Q_N(\cdot)$ also decreases, making $(1-Q_N(\cdot))^K$ closer to $1-KQ_N(\cdot)$. Consequently, the approximation of $\epsilon_{\text{WET-EHP}}^{i}$ in \eqref{ApproximateWET} is more accurate, leading to a more precise closed-form expression for general IDET. Additionally, the approximated closed-form for special IDET outage probability is more accurate when $W$ is larger, \textit{i.e.}, $W=6$. This is because the correlation parameter among different ports $\mu$ is smaller with a higher $W$, thereby making it more consistent with the conditions suitable for the approximation in \eqref{approximateIDET}. 
	
	\begin{figure*}[htbp]
		\centering
		\begin{minipage}[t]{0.3\textwidth}
			\centering
			\includegraphics[width=\textwidth]{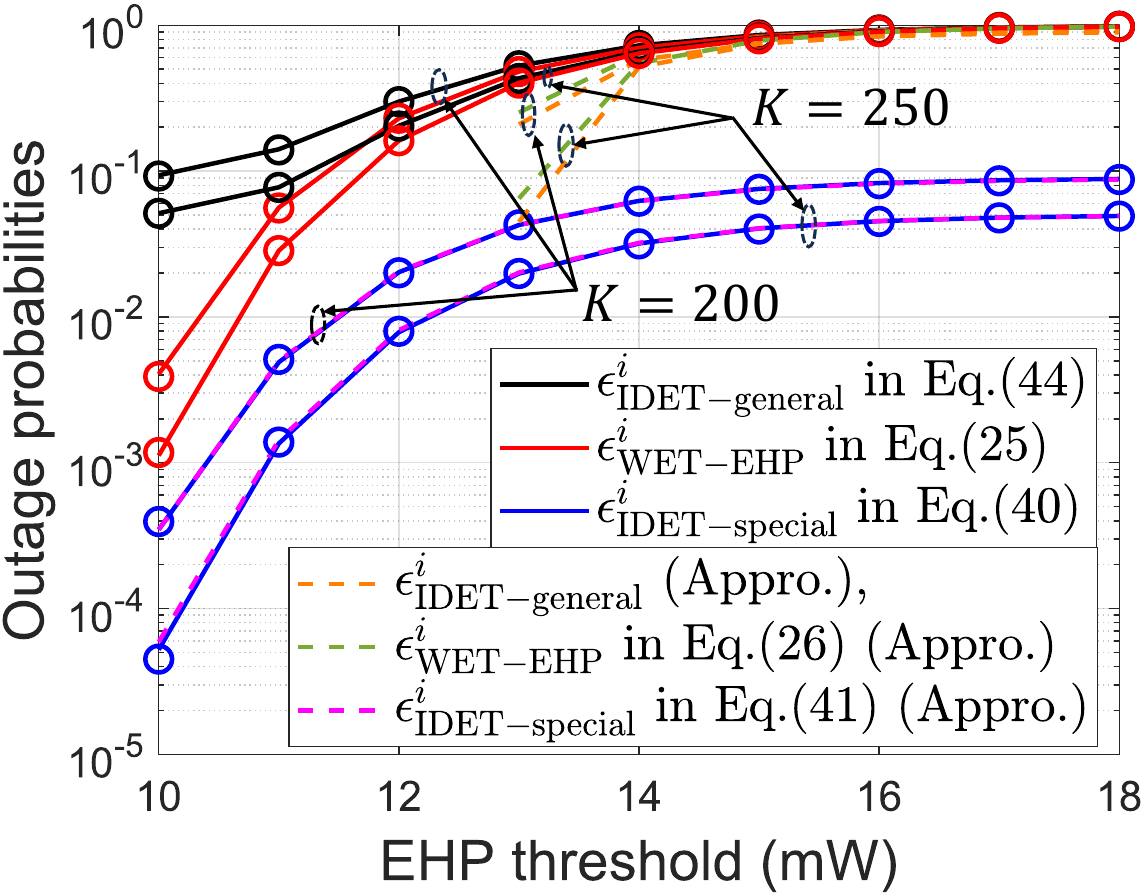}
			\caption{Outage probabilities versus $Q_{\text{th}}$.}
			\label{FigQ}
		\end{minipage}\hspace{0.01\linewidth}
		\begin{minipage}[t]{0.3\textwidth}
			\centering
			\includegraphics[width=\textwidth]{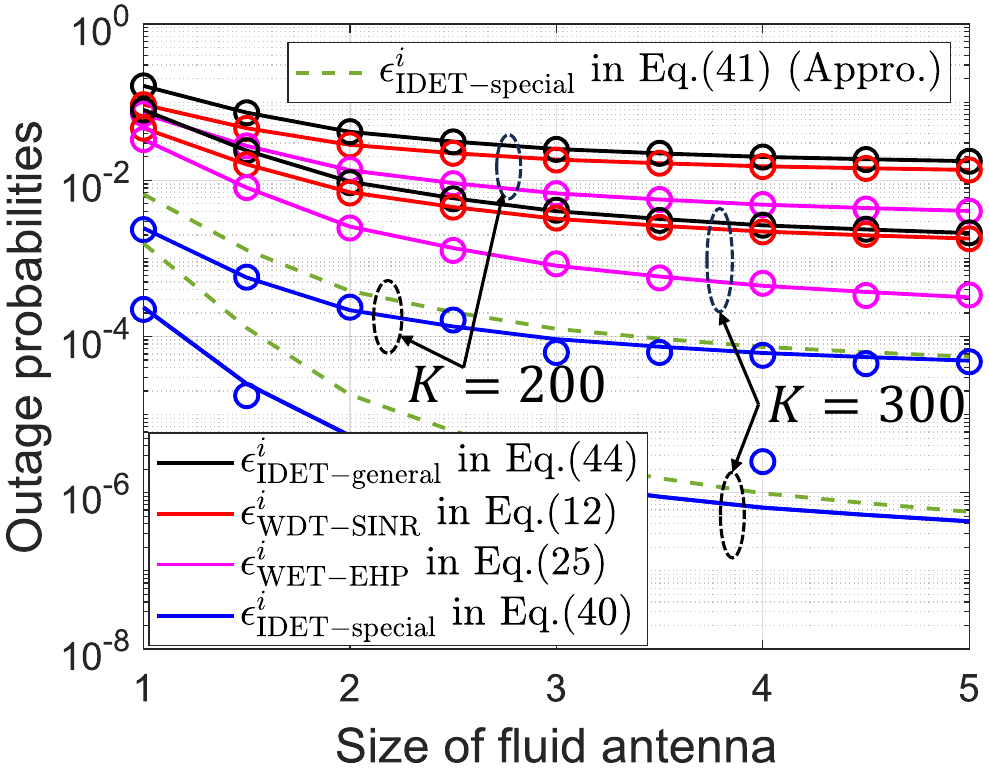}
			\caption{Outage probabilities versus $W$; $\gamma_{\text{th}}=2$ dB.}
			\label{FigW}
		\end{minipage}
		\begin{minipage}[t]{0.3\textwidth}
			\centering
			\includegraphics[width=\textwidth]{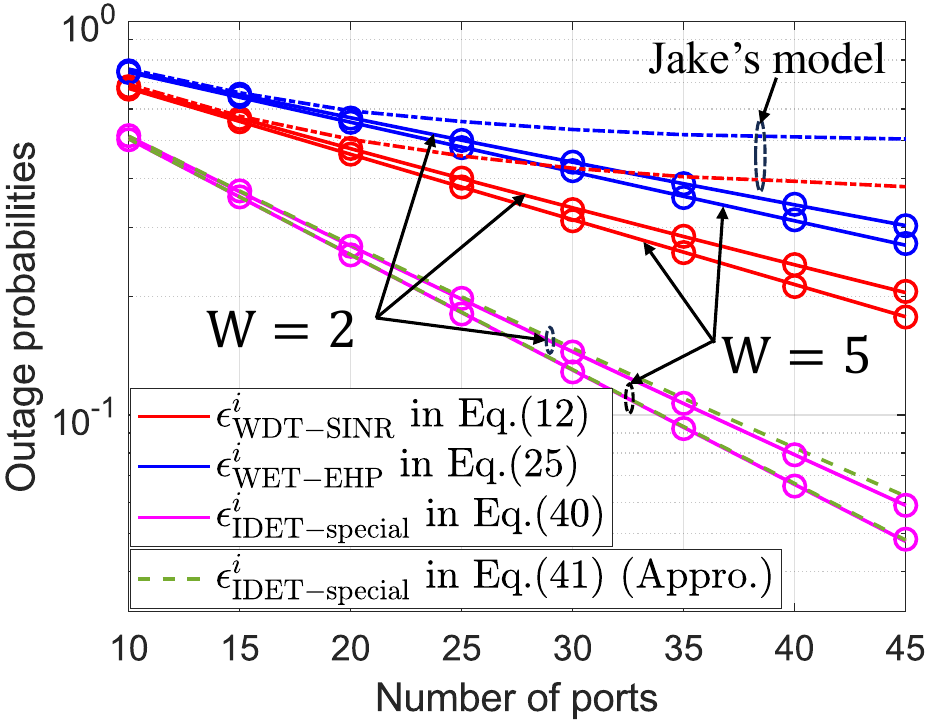}
			\caption{Outage probabilities versus $K$; $\gamma_{\text{th}}=1$ dB.}
			\label{FigK}
		\end{minipage}\hspace{0.01\linewidth}

		\begin{minipage}[t]{0.3\textwidth}
			\centering
			\includegraphics[width=\textwidth]{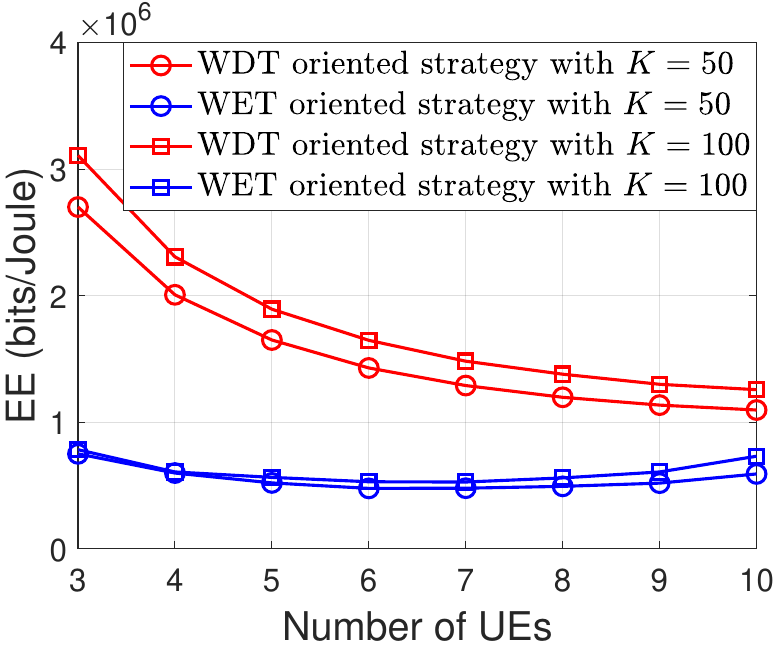}
			\caption{EE versus $N$; $P=1$ W, $W=5$, $B=1$ MHz, $P_C=0.5$ W, and $d=5$ m.}
			\label{FigEE}
		\end{minipage}\hspace{0.01\linewidth}
		\begin{minipage}[t]{0.3\textwidth}
			\centering
			\includegraphics[width=\textwidth]{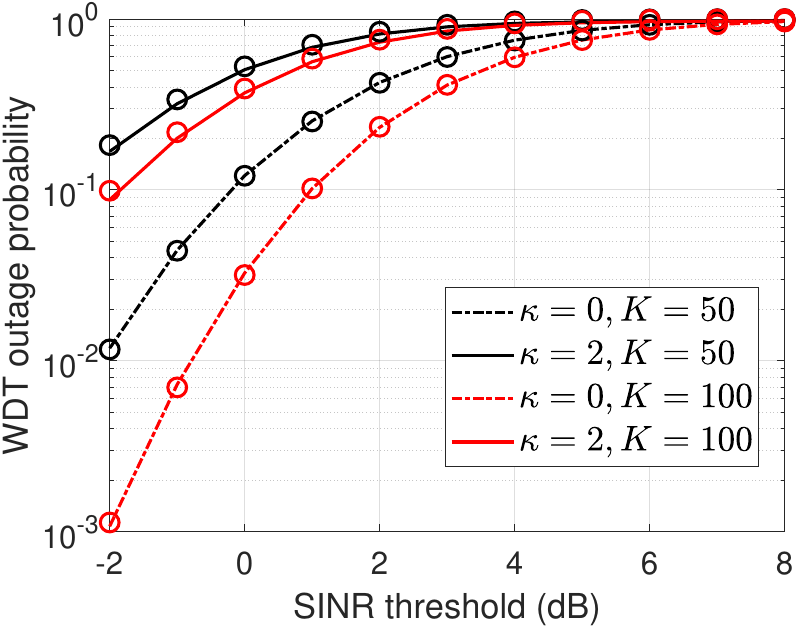}
			\caption{WDT outage probability with WDT oriented port strategy under Rician channel versus $\gamma_{\text{th}}$; $W=1$ and $N=5$.}
			\label{FigSINRRician}
		\end{minipage}\hspace{0.01\linewidth}
		\begin{minipage}[t]{0.3\textwidth}
			\centering
			\includegraphics[width=\textwidth]{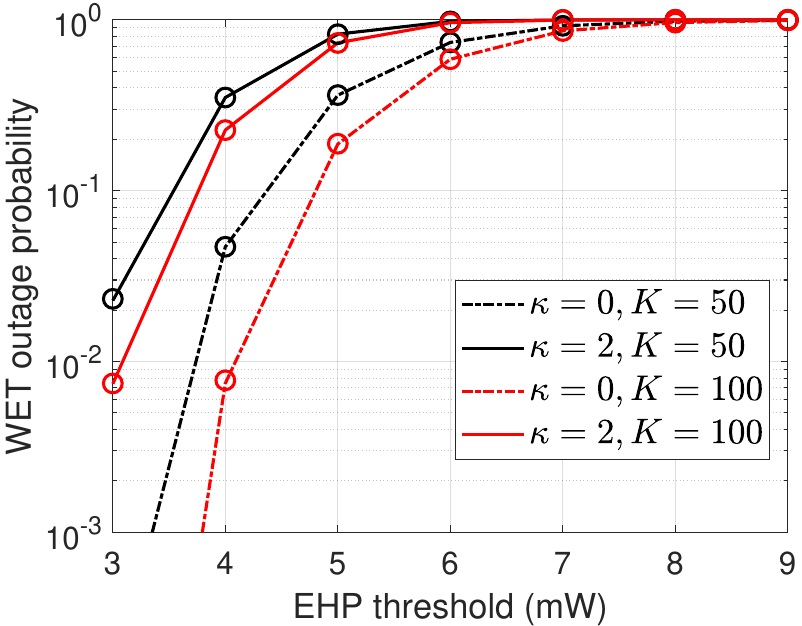}
			\caption{WET outage probability with WET oriented port strategy under Rician channel versus $Q_{\text{th}}$; $W=1$ and $N=5$.}
			\label{FigEHPRician}
		\end{minipage}\hspace{0.01\linewidth}
	\end{figure*}	
	\cref{FigQ} depicts the WET outage probability with the WET oriented port selection strategy, as well as the IDET outage probabilities versus $Q_{\text{th}}$ by conceiving different $K$. As depicted in \cref{FigQ}, the derived approximated closed-forms for WET outage probability in \eqref{ApproximateWET} and for general IDET outage probability in \eqref{IDETdefine2} closely approximate the exact value when the EHP threshold is relatively high, \textit{i.e.}, $Q_{\text{th}} = 14$ mW. Furthermore, we observe from \cref{FigQ} that a larger port number $K$ concurrently improves both the WET and IDET performance. This is because a larger $K$ increases the likelihood of selecting an optimal port for having a stronger wireless channel gain, which enhances the received signal strength for IDET.
	
	To explore the effect of fluid antenna size $W$ on the outage probabilities of the FAMA-assisted IDET system, we provide the results in Fig.~\ref{FigW} by considering different $K$. As expected, \cref{FigW} illustrates that the outage probabilities decrease with the fluid antenna size $W$.  Besides, when $W$ is small, \textit{i.e.}, $W=1$, the outage probability becomes unacceptably large, indicating that FAMA is not efficient anymore. It is also seen that when the fluid antenna size $W$ is small, the performance of IDET improves significantly with $W$ (from $W=1$ to $W=3$). However, when $W$ becomes larger, the performance improvement of the IDET system tends to flatten when we increase $W$ (from $W=3$ to $W=5$). This indicates that an appropriate fluid antenna size $W$ should be determined by considering the performance improvement as well as the hardware size. Additionally, as $W$ gradually increases, the approximation of the special IDET outage probability in \eqref{approximateIDET} tends to approach its exact value since the correlation among different ports is decreasing by increasing $W$. Also, a larger port number $K$ results in a larger approximation error. This is because by increasing $K$, the spatial correlation among different ports gradually increases, which results in a weaker independence assumption.
	
To further evaluate the effect of port number $K$ on the performance of the FAMA-assisted IDET system, Fig.~\ref{FigK} depicts the outage probabilities versus $K$ by considering two different fluid antenna sizes,  $W$. The performance of the proposed system improves with a larger $K$ or $W$.  However, observe from Fig.~\ref{FigK} that  the
outage probabilities continue to drop even if $K$ is very large, which is different from the results from the accurate Jake’s model in  \cite{AnalyticalApproximation}, where the outage probabilities saturate with a larger $K$. This is because for the sake of the tractability of theoretical analysis, we conceive a simplified channel model, which results in a lower correlation between ports and achieves abnormally better performance. Additionally, we have included a comparison between the simplified channel model used in this paper and the accurate Jake’s model with $W=5$. From the figure, it can be observed that, when $K$ is not large (\emph{i.e.,} $K\leq 20$), the expressions derived from the simplified channel model can effectively characterize the performance of the FAMA-assisted IDET system.
	
	\cref{FigEE} investigates the EE of the system versus UE number $N$ as well as different port number $K$ by considering different port selection strategies. It is observed that under the WDT oriented strategy, EE shows a gradual decline as $N$ increases, while under the WET oriented strategy, EE is convex with respect to $N$.  Besides, when $N$ is relatively small, the EE under the WDT oriented strategy is significantly higher than that under the WET oriented strategy. This is because the energy collected by the system is lower, which makes the sum rate a more critical determining factor. Furthermore, EE is improved as $K$ increases.
	
	\cref{FigSINRRician} and \cref{FigEHPRician} plot the WDT and WET outage probabilities by considering different port selection strategies under Rician channels. Observe from these two figures that  the presence of the LoS component negatively impacts the system performance.  This is because when we increase the Rician factor, the dominance of the LoS component increases, thereby reducing the contribution of the NLoS component. Therefore, the variation of the channel gain decreases, resulting in a lower probability that the FA switches to the port where the interference suffers from deep fade or EHP achieves its peak. Moreover, when we increase the number of ports $K$,  outage probabilities for both WDT and WET decrease, indicating that the FAMA system can still function well under the Rician channel model.

	\section{Conclusion}\label{section6}
	In this paper, the novel FAMA-assisted IDET system has been investigated by employing the PS approach, where each UE is equipped with a single FA to mitigate multiuser interference or enhance the receive signal strength for energy harvesting by dynamically switching the antenna ports. Specifically, the exact outage probabilities for WDT and WET with different port selection strategies were derived and approximated in closed forms under the Rayleigh channel model, while the special and general IDET outage probabilities were also analyzed to evaluate the IDET performance of the FAMA system. Additionally, multiplexing gains for WDT, WET, and IDET were provided to explore potential system performance. Furthermore, we extend the analysis to the Rician channel model, offering a more general analytical framework. Simulation results validated our theoretical analysis and evaluated the performance of the proposed system, providing insights into the FAMA-assisted IDET system design, \textit{e.g.}, both WDT and WET exhibit poor performance by considering the undesired strategies.
	
	\appendices
	
	\section{Proof of Theorem 1}\label{proofth1}
	\setcounter{equation}{0}
	\renewcommand{\theequation}{A.\arabic{equation}}
	
	As $\gamma_{\text{th}}$ increases, the value $\left[ \cdots\right]^K$ in \eqref{ExactWDTP} gradually decreases, indicating that binomial approximation can be relied upon. Then the exact WDT outage probability with the WDT oriented strategy for maximizing the SINR in \eqref{ExactWDTP} can be approximated as 
	\begin{equation}
		\begin{aligned}
			\epsilon_\text{{WDT}}^i&\approx\int_{0}^{\infty}\int_{0}^{\infty}\frac{\tilde{r}^{N-2}e^{-\frac{r+\tilde{r}}{2}}}{2^{N}\Gamma(N-1)}\\
			&\qquad\qquad\times\left[1-K\left( 1-Q_N(\cdot,\cdot+\cdots)\right) \right]drd\tilde{r}\\
			&=\left( 1-K\right) \beta_1+K\beta_2-K\beta_3	,
		\end{aligned}
	\end{equation}
	where $\beta_1$ and $\beta_2$ are defined in \cite[eq. (75)]{sfama} and \cite[eq. (76)]{sfama}, respectively, given by
	\begin{equation}
		\beta_1=\int_{0}^{\infty}\int_{0}^{\infty}\frac{\tilde{r}^{N-2}e^{-\frac{r+\tilde{r}}{2}}}{2^{N}\Gamma(N-1)}drd\tilde{r}=1,
	\end{equation}
	and
	\begin{equation}
		\begin{aligned}
			\beta_2&=\int_{0}^{\infty}\textbf{}\int_{0}^{\infty}\frac{\tilde{r}^{N-2}e^{-\frac{r+\tilde{r}}{2}}}{2^{N}\Gamma(N-1)}\times\\
			&\quad\quad Q_{N-1}\left(\frac{\mu\sqrt{\frac{\gamma_{\text{th}}\tilde{r}}{\gamma_{\text{th}}+1}}}{\sqrt{1-\mu^2}},\frac{\mu\sqrt{\frac{r}{\gamma_{\text{th}}+1}}}{\sqrt{1-\mu^2}}\right)drd\tilde{r}\\
			&=1-\left(\frac{\mu^2}{\gamma_{\text{th}}+1}\right)^{N-1}.
		\end{aligned}
	\end{equation}
	
	Afterwards, $\beta_3$ can be formulated as \eqref{beta3} (see top of next page). 
	\begin{figure*}[]
		% ensure that we have normalsize text
		\normalsize
		% Store the current equation number.
		\setcounter{MYtempeqncnt}{\value{equation}}
		\setcounter{equation}{3}
		\begin{equation}\label{beta3}
			\begin{aligned}
				\beta_3&\hspace{.5mm}=\left(\frac{1}{\gamma_{\text{th}}+1}\right)^{N-1}\sum_{k=0}^{N-2}\sum_{j=0}^{N-k-2}(\gamma_{\text{th}}+1)^k\gamma_{\text{th}}^{\frac{j-k}{2}}\frac{(N-(j+k)-1)_j}{j!}\times\\
				&\int_{0}^{\infty}\int_{0}^{\infty}\frac{\tilde{r}^{N-2}}{2^{N}\Gamma(N-1)}e^{-\frac{r+\tilde{r}}{2}}\exp{\left[-\frac{\mu^2}{2(1-\mu^2)}\left(\frac{\gamma\tilde{r}+r}{\gamma+1}\right)\right]}\left(\frac{r}{\tilde{r}}\right)^{\frac{j+k}{2}}I_{j+k}\left(\frac{\mu^2}{1-\mu^2}\frac{\sqrt{\gamma_{\text{th}} r \tilde{r}}}{\gamma_{\text{th}}+1}\right)drd\tilde{r}\\
				&\hspace{.5mm}=\left(\frac{1}{\gamma_{\text{th}}+1}\right)^{N-1}\sum_{k=0}^{N-2}\sum_{j=0}^{N-k-2}(\gamma_{\text{th}}+1)^k\gamma_{\text{th}}^{\frac{j-k}{2}}\frac{(N-(j+k)-1)_j}{j!}\int_{0}^{\infty}\frac{\tilde{r}^{N-2-\frac{j+k}{2}}}{2^{N}\Gamma(N-1)}\exp{\left[-\frac{1-\mu^2+\gamma_{\text{th}}}{2(1-\mu^2)(\gamma_{\text{th}}+1)}\right]\tilde{r}}d\tilde{r}\\
				&\int_{0}^{\infty}r^{\frac{j+k}{2}}\exp{\left[-\frac{(1-\mu^2)\gamma_{\text{th}}+1}{2(1-\mu^2)(\gamma_{\text{th}}+1)}r\right]}I_{j+k}\left(\frac{\mu^2}{1-\mu^2}\frac{\sqrt{\gamma_{\text{th}}\tilde{r}}}{\gamma_{\text{th}}+1}\sqrt{r}\right)dr\\
				&\overset{(a)}{=}\left(\frac{1}{\gamma_{\text{th}}+1}\right)^{N-1}\sum_{k=0}^{N-2}\sum_{j=0}^{N-k-2}(\gamma_{\text{th}}+1)^k\gamma_{\text{th}}^{\frac{j-k}{2}}\frac{(N-(j+k)-1)_j}{j!}\int_{0}^{\infty}\frac{\tilde{r}^{N-2-\frac{j+k}{2}}}{2^{N}\Gamma(N-1)}\exp{\left[-\frac{1-\mu^2+\gamma_{\text{th}}}{2(1-\mu^2)(\gamma_{\text{th}}+1)}\right]\tilde{r}}d\tilde{r}\\\\
				&\frac{2\left(1-\mu^2\right)\left(\gamma+1\right)}{\mu^2\sqrt{\gamma_{\text{th}}\tilde{r}}}e^{\frac{\mu^4\gamma_{\text{th}}\tilde{r}}{4\left(1-\mu^2\right)\left(\gamma_{\text{th}}+1\right)\left[\left(1-\mu^2\right)\gamma_{\text{th}}+1\right]}}\left[\frac{(1-\mu^2)\gamma_{\text{th}}+1}{2\left(1-\mu^2\right)\left(\gamma_{\text{th}}+1\right)}\right]^{-\frac{j+k+1}{2}}M_{-\frac{j+k+1}{2},\frac{j+k}{2}}\left(\frac{\mu^4\gamma_{\text{th}}\left[\left(1-\mu^2\right)\gamma_{\text{th}}+1\right]^{-1}\tilde{r}}{2\left(1-\mu^2\right)\left(\gamma_{\text{th}}+1\right)}\right)\\
				&\overset{(b)}{=}\left[\frac{\mu^4\gamma_{\text{th}}}{\left(1-\mu^2\right)\gamma_{\text{th}}+1}\right]^{\frac{j+k+1}{2}}  \frac{2\left(1-\mu^2\right)\left(\gamma_{\text{th}}+1\right)}{\mu^2\sqrt{\gamma_{\text{th}}}\left[\left(1-\mu^2\right)\gamma_{\text{th}}+1\right]^{\frac{j+k+1}{2}}}\int_{0}^{\infty}\frac{\tilde{r}^{N-2}}{2^{N}\Gamma(N-1)}e^{-\left(\frac{1-\mu^2+\gamma_{\text{th}}}{2\left(1-\mu^2\right)\left(\gamma_{\text{th}}+1\right)}+\frac{\mu^4\gamma_{\text{th}}\left[\left(1-\mu^2\right)\gamma_{\text{th}}+1\right]^{-1}}{4\left(1-\mu^2\right)\left(\gamma_{\text{th}}+1\right)}\right)\tilde{r}}d\tilde{r}\\
				&\overset{(c)}{=}\left[\frac{2\gamma_{\text{th}}\left(1-\mu^2\right)+1}{2\gamma_{\text{th}}^2+\left(3-\mu^2\right)\gamma_{\text{th}}+1}\right]^{N-1}\left(1-\mu^2\right)\sum_{k=0}^{N-2}\sum_{j=0}^{N-k-2}(\gamma_{\text{th}}+1)^{k+1}\gamma_{\text{th}}^{j}\frac{\left(N-\left(j+k\right)-1\right)_j}{j!}\frac{\mu^{2(j+k)}}{\left[\left(1-\mu^2\right)\gamma_{\text{th}}+1\right]^{j+k+1}}
			\end{aligned}
		\end{equation}
		\hrulefill
		% The spacer can be tweaked to stop underfull vboxes.
	\end{figure*}	
	By applying the results in \cite[eq. (6.643.2)]{handbook}, the outer integral on $r$ is evaluated and $(a)$ is derived, where $M_{u,v}(\cdot)$ is the Whittaker function. To evaluate the inner integral on $\tilde{r}$, we exploit an important connection between Whittaker functions $M_{u,v}(\cdot)$ and generalized hypergeometric functions $_{a}F_b(\cdot;\cdot;\cdot)$ established in \cite[eq. (9.220.2)]{handbook}. By leveraging the expansion for confluent hypergeometric function, $(b)$ is obtained after simplifications. Then, by applying the result in \cite[eq. (3.326.2)]{handbook} and after simplifications, $(c)$ is obtained. As a result, the operation $[\cdot]^{+}$ is adopted to guarantee positivity of the approximated expression, which proves Theorem \ref{th1}.
	
	\renewcommand{\theequation}{B.\arabic{equation}}
	\setcounter{equation}{0}
	\section{Proof of Theorem 2}\label{proofth2}
	First of all, we define $\tilde{r}_1=(x_0^{(i,i)})^2+(y_0^{(i,i)})^2$, $\tilde{r}_2=\sum_{m\neq i}^N(x_0^{(m,i)})^2+(y_0^{(m,i)})^2$, which are central chi-square distributed with $2$ degrees of freedom and  $2(N-1)$ degrees of freedom, respectively. Hence, the probability density function (PDF) of $\tilde{r}_1$ and $\tilde{r}_2$ is expressed as
	\begin{align}
		f_{\tilde{r}_1}(r_1)&=\frac{1}{2}\exp\left(-\frac{r_1}{2}\right),\label{pdfr1}\\
		f_{\tilde{r}_2}(r_2)&=\frac{r_2^{N-2}\exp\left(-\frac{r_2}{2}\right)}{2^{N-1}\Gamma(N-1)}.\label{pdfr2}
	\end{align}
	Then, we define $\mathcal{A}=\arg\max_{n\neq {k^{\ast}}} \frac{X_n}{Y_n}$. Now, our objective is to derive the outage probability as $\text{Prob}(Y_{k^{\ast}}\mathcal{A}<X_{k^{\ast}}<\widehat{Q}_{\text{th}}-Y_{k^{\ast}})$. In order to derive the WET outage probability with the WDT oriented strategy, the distribution of $\mathcal{A}$ and $X_{k^{\ast}}$ should firstly be formulated, respectively. The CDF of $\mathcal{A}$ conditioned on $\tilde{r}_1$ and $\tilde{r}_2$ is given by \cite[eq. (62)]{sfama}
	\begin{equation}
		\begin{aligned}
			&F_{\mathcal{A}\vert \tilde{r}_1,\tilde{r}_2}(z)=\text{Prob}\left(\mathcal{A}=\arg\max_{n\neq k^{\ast}}\frac{X_n}{Y_n}<z\right)\\
			&=\left[1-\frac{1}{2}\int_{0}^{\infty}Q_1\left(\sqrt{\frac{\mu^2r_1}{1-\mu^2}},\sqrt{zy}\right)\left(\frac{1-\mu^2y}{\mu^2r_2}\right)^{\frac{N-2}{2}}\right.\\
			&\enspace\left.\times\exp\left(-\frac{y+\frac{\mu^2}{1-\mu^2}r_2}{2}\right)I_{N-2}\left(\sqrt{\frac{\mu^2}{1-\mu^2}r_2y}\right)dy\right]^{K-1}.
		\end{aligned}
	\end{equation}
	By differentiating the CDF of $\mathcal{A}$, the PDF of $\mathcal{A}$ conditioned on $\tilde{r}_1$ and $\tilde{r}_2$ is obtained as
	\begin{equation}\label{f_A}
		\begin{aligned}
			&f_{\mathcal{A}\vert \tilde{r}_1,\tilde{r}_2}(z)=\frac{\partial F_{\mathcal{A}\vert\tilde{r}_1,\tilde{r}_2}(z)}{\partial z}\\
			&=(K-1)\left[1-\frac{1}{2}\int_{0}^{\infty}Q_1\left(\sqrt{\frac{\mu^2}{1-\mu^2}r_1},\sqrt{zy}\right)\times\right.\\
			&\enspace\enspace\enspace \left.e^{-\frac{y+\frac{\mu^2}{1-\mu^2}r_2}{2}}I_{N-2}\left(\sqrt{\frac{\mu^2}{1-\mu^2}}\sqrt{r_2y}\right)dy\right]^{K-2} \\
			&\enspace\enspace\times\int_{0}^{\infty}\frac{y}{2}e^{-\frac{\frac{\mu^2}{1-\mu^2}r_1+zy}{2}}I_0\left(\sqrt{\frac{\mu^2}{1-\mu^2}r_1zy}\right)\times\\ &\enspace\enspace\enspace\frac{1}{2}\left(\frac{y}{\frac{\mu^2}{1-\mu^2}r_2}\right)^{\frac{N-2}{2}} e^{-\frac{y+\frac{\mu^2}{1-\mu^2}r_2}{2}}I_{N-2}\left(\sqrt{\frac{\mu^2}{1-\mu^2}r_2y}\right)dy.
		\end{aligned}
	\end{equation}
	By applying the results in \cite[eq. (59)]{sfama}, the CDF of $X_{k^{\ast}}$ conditioned on $\tilde{r}_1$ is obtained as 
	\begin{equation}\label{CDFXi}
		F_{X_{k^{\ast}}\vert \tilde{r}_1}(t)=\left[1-Q_1\left(\sqrt{\frac{\mu^2}{1-\mu^2}r_1},\sqrt{t}\right)\right].
	\end{equation}
	Further, by applying the results in \cite[eq. (61)]{sfama}, the PDF of $Y_{k^{\ast}}$ conditioned on $\tilde{r}_2$ is obtained as by substituting the variable $K=1$ is obtained as
	\begin{multline}\label{PDFYi}
		f_{Y_{k^{\ast}}\vert \tilde{r}_2}(y)=\frac{1}{2}\left(\frac{\left(1-\mu^2\right)y}{\mu^2r_2}\right)^{\frac{N-2}{2}} \exp\left(-\frac{y+\frac{\mu^2r_2}{1-\mu^2}}{2} \right) \\
		\times I_{N-2}\left(\sqrt{\frac{\mu^2r_2y}{1-\mu^2}}\right).
	\end{multline}
	As a consequence, the probability $\text{Prob}(Y_{k^{\ast}}\mathcal{A}<X_{k^{\ast}}<\hat{Q}_{\text{th}}-Y_{k^{\ast}})$ is then re-formulated as \eqref{processconditionalWET}, 
	\begin{figure*}[!t]
		\renewcommand{\theequation}{B.\arabic{equation}}
		\setcounter{equation}{6}
		\begin{equation}\label{processconditionalWET}
			\begin{aligned}
				&\text{Prob}\left(Y_{k^{\ast}} \max_{n\neq {k^{\ast}}} \frac{X_n}{Y_n}<X_{k^{\ast}}<\widehat{Q}_{\text{th}}-Y_{k^{\ast}}\right)=\text{Prob}(Y_{k^{\ast}}\mathcal{A}<X_{k^{\ast}}<\widehat{Q}_{\text{th}}-Y_{k^{\ast}})\\
				&\overset{(a)}{=}\int_{z=0}^{\infty}\int_{y=0}^{\frac{\widehat{Q}_{\text{th}}}{1+z}}\text{Prob}(Y_{k^{\ast}}\mathcal{A}<X_{k^{\ast}}<\widehat{Q}_{\text{th}}-Y_{k^{\ast}}\vert Y_{k^{\ast}},\mathcal{A})f_{Y_{k^{\ast}}}(y)f_{\mathcal{A}}(z)dydz\\
				&\overset{(b)}{=}\int_{r_1=0}^{\infty}\int_{r_2=0}^{\infty}\int_{z=0}^{\infty}\int_{y=0}^{\frac{\widehat{Q}_{\text{th}}}{1+z}}\underbrace{\frac{r_2^{N-2}\exp\left(-\frac{r_1+r_2}{2}\right)}{2^{N}\Gamma(N-1)}}_{\text{	$f_{\tilde{r}_1}(r_1)f_{\tilde{r}_2}(r_2)$}}\left[F_{X_{k^{\ast}}\vert\tilde{r}_1}(\widehat{Q}_{\text{th}}-y)-F_{X_{k^{\ast}}\vert\tilde{r}_1 }(yz)\right]f_{Y_{k^{\ast}}\vert\tilde{r}_2}(y)f_{\mathcal{A}\vert\tilde{r}_1,\tilde{r}_2}(z)dydzdr_1dr_2\\
			\end{aligned}
		\end{equation}
		\hrulefill
	\end{figure*}
	where $ (a) $ accounts for the fact that $X_{k^{\ast}}$, $Y_{k^{\ast}}$ and $\mathcal{A}$ are independent when conditioned on $\tilde{r}_1$ and $\tilde{r}_2$, $(b)$ is derived with the results of PDF of $\tilde{r}_1$ and $\tilde{r}_2$. Finally, the desired result \eqref{WETMSC} is obtained by applying the results in \eqref{f_A}, \eqref{CDFXi} and \eqref{PDFYi}, which finally proves Theorem \ref{th2}.
	
	\section{Proof of Lemma 1}\label{prooflemma1}
	\setcounter{equation}{0}
	\renewcommand{\theequation}{C.\arabic{equation}}
	When $\mu=0$, $X_k$ and $Y_k$ follow central chi-squared distributions with $2$ degrees of freedom and $2(N-1)$ degrees of freedom, respectively. Thus, we have
	\begin{align}
		\widehat{X}_k&=\lim\limits_{\mu\rightarrow0}{X_k}=\left(x_k^{(i,i)}\right)^2+\left(y_k^{(i,i)}\right)^2,\label{approximateX_k}\\
		\widehat{Y}_k&=\lim\limits_{\mu\rightarrow0}{Y_k}=\sum_{m\neq i }^{N}\left[\left(x_k^{(m,i)}\right)^2+\left(y_k^{(m,i)}\right)^2\right].\label{approximateY_k}
	\end{align}
	Then, the $\widehat{X}_{k}$ is exponentially distributed with PDF 
	\begin{equation}
		f_{\widehat{X}_{k}}\left(x\right)=e^{-x/2},
	\end{equation}
	and 	$\widehat{Y}_{k}$ is a central Chi-square distribution with PDF
	\begin{equation}\label{approximateYkmu=0}
		f_{\widehat{Y}_{k}}\left(y\right)=\frac{1}{2^{N-1}\Gamma\left(N-1\right)}y^{N-2}e^{-y/2}.
	\end{equation}
	After that, we use the transformations: $V_1=\widehat{X}_{k}+\widehat{Y}_{k}$ and $V_2=\widehat{X}_{k}/\widehat{Y}_{k}$. Since $\widehat{X}_{k}$ and $\widehat{Y}_{k}$ are independent, $V_1$ is a central Chi-square distribution with the PDF
	\begin{equation}
		f_{V_1}\left(v_1\right) = \frac{1}{2^{N}\Gamma\left(N\right)}{v_1}^{N-1}e^{-v_1/2},
	\end{equation}
	and $V_2$ is a Beta prime distribution with the PDF
	\begin{equation}
		f_{V_2}\left(v_2\right) = \frac{N-1}{\left(1+v_2\right)^N}.
	\end{equation}
	Then, we have $\widehat{X}_{k}=\frac{V_1V_2}{1+V_2}$ and $\widehat{Y}_{k}=\frac{V_1}{1+V_2}$. By formulating the absolute value of the Jacobian determinant as $\left\vert J\right\vert = \frac{V_1}{\left(1+V_2\right)^2}$, the joint PDF of $V_1$ and $V_2$ is
	\begin{equation}
		f_{V_1,V_2}\left(v_1,v_2\right)=f_{\widehat{X}_{k},\widehat{Y}_{k}}\left(x,y\right)\cdot\left\vert J\right\vert.
	\end{equation}
	Finally, we have the following results by substituting the PDF of $\widehat{X}_{k}$, $\widehat{Y}_{k}$ as
	\begin{equation}
		f_{V_1,V_2}\left(v_1,v_2\right)=\underbrace{\frac{v_1^{N-1}e^{-v_1/2}}{2^N\Gamma\left(N\right)}}_{\text{PDF of } V_1}\cdot\underbrace{\frac{N-1}{\left(1+v_2\right)^N}}_{\text{PDF of }V_2}.
	\end{equation}
	Since the joint PDF factorizes into the product of marginal PDFs, $V_ 1$ and $V_2$ are independent. Therefore, the proof of Lemma \ref{lemma2} ends.
	
	\renewcommand{\theequation}{D.\arabic{equation}}
	\setcounter{equation}{0}
	\section{Proof of Theorem 3}\label{proofth3}
	When $\widehat{Q}_{\text{th}}$ becomes larger or $N$ gets smaller, the Marcum-$Q$ function $Q_N\left(\sqrt{\frac{\mu^2}{1-\mu^2}r},\sqrt{\widehat{Q}_{\text{th}}}\right)$ goes to zero. Hence, the binomial approximation can be applied as
	\begin{multline}\label{biWET}
		\left[1-Q_N\left(\sqrt{\frac{\mu^2r}{1-\mu^2}},\sqrt{\widehat{Q}_{\text{th}}}\right)\right]^K\\
		\approx1-KQ_N\left(\sqrt{\frac{\mu^2r}{1-\mu^2}},\sqrt{\widehat{Q}_{\text{th}}}\right).
	\end{multline}
	Then the exact WET outage probability with the WET oriented strategy for maximizing the EHP in \eqref{exactWETP} is approximated as
	\begin{equation}\label{Biin}
		\begin{aligned}
			&\epsilon^i_{\text{WET-EHP}}\\
			&\approx\int_{0}^{\infty}\frac{r^{N-1}e^{-\frac{r}{2}}}{2^N\Gamma(N)}\left[1-KQ_N\left(\sqrt{\frac{\mu^2r}{1-\mu^2}},\sqrt{\widehat{Q}_{\text{th}}}\right)\right]dr\\
			&=1-K\int_{0}^{\infty}\frac{r^{N-1}e^{-\frac{r}{2}}}{2^N\Gamma(N)}Q_N\left(\sqrt{\frac{\mu^2r}{1-\mu^2}},\sqrt{\widehat{Q}_{\text{th}}}\right)dr.
		\end{aligned}
	\end{equation}
	By applying the results in \cite{Sofotasios}, a closed-form solution for the following type of integral is available, i.e.,
	\begin{equation}
		\mathcal{F}(k,m,a,b,p)=\int_0^{\infty}x^{k-1}Q_m(a\sqrt{x},b)e^{-px}dx.
	\end{equation}
	By using the result in \cite[eq. (12)]{Sofotasios} to evaluate the integral $\mathcal{F}\left(N, N, \sqrt{\frac{\mu^2}{1-\mu^2}}, \sqrt{\widehat{Q}_{\text{th}}}, \frac{1}{2}\right)$, and adopting the operation $[\cdot]^{+}$ to guarantee positivity of the approximated expression, the approximated WET outage probability with the WET oriented strategy is obtained in \eqref{ApproximateWET}, which proves Theorem \ref{th3}.
	
	\renewcommand{\theequation}{E.\arabic{equation}}
	\section{Proof of Theorem 4}\label{proofth4}
	\setcounter{equation}{0}
	We define $\mathcal{B}=\arg\max_{n\neq {k^{\ast}}} X_n+Y_n$, where $k^{\ast}$ is the optimal port. Now, our objective is to derive the outage probability as $\text{Prob}(\mathcal{B}-Y_{k^{\ast}} <X_{k^{\ast}}<\gamma Y_{k^{\ast}}) $, where the PDF of $X_{k^{\ast}}$ and the distribution of $\mathcal{B}$ should be firstly formulated. By applying the results in \cite[eq. (58)]{sfama},  the unconditioned PDF of $X_{k^{\ast}}$  by substituting $K=1$ is given by
	\begin{equation}\label{PDFXi}
		f_{X_{k^{\ast}}}(t)=\int_{0}^{\infty}\underbrace{\frac{\exp{\left(-\frac{r_1}{2}\right)}}{2}}_{\text{$f_{\tilde{r}_1}(r_1)$}}\frac{e^{-\frac{\frac{\mu^2}{1-\mu^2}r_1+t}{2}}}{2}I_0\left(\sqrt{\frac{\mu^2r_1t}{1-\mu^2}}\right)dr_1.
	\end{equation}
	Similarly, the PDF $f_{\mathcal{B}}(z)$ of $\mathcal{B}$ conditioned on $\tilde{r}_1$ and $\tilde{r}_2$ can be formulated using \cite[eq. (26)]{lin}, while the CDF of $\mathcal{B}$ conditioned on $\tilde{r}_1$ and $\tilde{r}_2$ is formulated as \cite[eq. (19)]{lin}
	\begin{equation}\label{CDFB}
		F_{\mathcal{B}\vert \tilde{r}_1,\tilde{r}_2}(z)=\left[1-Q_N\left(\sqrt{\frac{\mu^2}{1-\mu^2}(r_1+r_2)},\sqrt{z}\right)\right]^{K-1}.
	\end{equation}
	
	Then, the probability $\text{Prob}(\mathcal{B}-Y_{k^{\ast}} <X_{k^{\ast}}<\gamma Y_{k^{\ast}})$ is evaluated by \eqref{processconditionalWDT} (see top of next page), 
	\begin{figure*}
		\renewcommand{\theequation}{E.\arabic{equation}}
		\setcounter{equation}{2}
		\begin{equation}\label{processconditionalWDT}
			\begin{aligned}
				&\text{Prob}(\mathcal{B}-Y_{k^{\ast}} <X_{k^{\ast}}<\gamma Y_{k^{\ast}})\\
				&\overset{(a)}{=}\int_{y=0}^{\infty}\int_{z=0}^{(1+\gamma_{\text{th}})y}\text{Prob}(\mathcal{B}-Y_{k^{\ast}} <X_{k^{\ast}}<\gamma Y_{k^{\ast}}\vert Y_{k^{\ast}},\mathcal{B})f_{Y_{k^{\ast}}}(y)f_{\mathcal{B}}(z)dydz\\
				&\overset{(b)}{=}\int_{r_1=0}^{\infty}\int_{r_2=0}^{\infty}\int_{y=0}^{\infty}\int_{z=0}^{(1+\gamma_{\text{th}})y}\underbrace{	\frac{r_2^{N-2}\exp\left(-\frac{r_1+r_2}{2}\right)}{2^{N}\Gamma(N-1)}}_{\text{$f_{\tilde{r}_1}(r_1)f_{\tilde{r}_2}(r_2)$}}\left[F_{X_{k^{\ast}}\vert \tilde{r}_1}(\gamma_{\text{th}} y)-F_{X_{k^{\ast}}\vert \tilde{r}_1}(z-y)\right]f_{Y_{k^{\ast}}\vert \tilde{r}_2}(y)f_{\mathcal{B}\vert \tilde{r}_1,\tilde{r}_2}(z)dydzdr_1dr_2\\
				&\overset{(c)}{=}\int_{r_1=0}^{\infty}\int_{r_2=0}^{\infty}\int_{y=0}^{\infty}\int_{t=0}^{\gamma_{\text{th}} y}\underbrace{	\frac{r_2^{N-2}\exp\left(-\frac{r_1+r_2}{2}\right)}{2^{N}\Gamma(N-1)}}_{\text{$f_{\tilde{r}_1}(r_1)f_{\tilde{r}_2}(r_2)$}}F_{\mathcal{B}\vert \tilde{r}_1,\tilde{r}_2}(t+y)f_{X_{k^{\ast}}}(t)f_{Y_{k^{\ast}}}(y)dtdydr_1dr_2
			\end{aligned}
		\end{equation}
		\hrulefill
	\end{figure*}
	where $ (a) $ accounts for the fact that $X_{k^{\ast}}$, $Y_{k^{\ast}}$ and $\mathcal{B}$ are independent when conditioned on $\tilde{r}_1$ and $\tilde{r}_2$, $(b)$ is derived by using the PDF of $\tilde{r}_1$ and $\tilde{r}_2$, $(c)$  is obtained by integrating $z$ by parts and subsequent simplification. Then, \eqref{WDTMHP} is achieved by applying the results in \eqref{PDFYi}, \eqref{PDFXi} and \eqref{CDFB}, which finally proves Theorem \ref{th4}.
		
\renewcommand{\theequation}{F.\arabic{equation}}
\setcounter{equation}{0}
\section{Proof of Theorem 5}\label{proofth5}
Note that $X_1, \dots ,X_K$ are all independent with each other when conditioned on $\tilde{r}_1$, which follows chi-square distribution with $2$ degrees of freedom.  Hence, we can get the joint PDF of $X_1, \dots ,X_K$ as \cite[eq. (58)]{sfama}
\begin{multline}\label{jointfX}
	f_{X_1,\dots,X_K}(t_1,\dots,t_K)=\int_{0}^{\infty}\underbrace{\frac{\exp{\left(-\frac{r_1}{2}\right)}}{2}}_{\text{$f_{\tilde{r}_1}(r_1)$}}\times\\
	\prod_{k=1}^{K}\frac{1}{2}e^{-\frac{t_k+\frac{\mu^2}{1-\mu^2}r_1}{2}}I_0\left(\sqrt{\frac{\mu^2}{1-\mu^2}}\sqrt{r_1t_k}\right)dr_1.
\end{multline}
Similarly, $Y_1, \dots ,Y_K$ are all independent with each other when conditioned on $\tilde{r}_2$, which follows chi-square distribution with $2(N-1)$ degrees of freedom. Then we can get the joint CDF of $Y_1, \dots ,Y_K$ as \cite[eq. (17)]{lin}
\begin{multline}\label{jointFY}
	F_{Y_1,\dots,Y_K}(t_1,\dots,t_K)=\int_{0}^{\infty}\underbrace{\frac{r_2^{N-2}\exp{\left(-\frac{r_2}{2}\right)}}{2^{N-1}\Gamma(N-1)}}_{\text{$f_{\tilde{r}_2}(r_2)$}}\times\\
	\prod_{k=1}^{K}\left[1-Q_{N-1}\left(\sqrt{\frac{\mu^2}{1-\mu^2}r_2},\sqrt{t_k}\right)\right]dr_2.
\end{multline}
Further, the special IDET outage probability $\text{Prob}\left(\text{SINR}_k^i < \gamma_{\text{th}}, Q_k^i < Q_{\text{th}}\right), \forall k= 1, 2, \dots, K$ is evaluated by \eqref{jointprocess} (see next page). Note that ($a$) accounts for the fact that $X_k$ and $Y_k$ $\left(k= 1,\dots, K\right)$ are independent when conditioned on $\tilde{r}_1$ and $\tilde{r}_2$, ($b$) is derived according to \eqref{jointfX} and \eqref{jointFY}, and ($c$) moves the integration over $X_k$ inside the product. Finally, \eqref{finalIDET} is achieved and then the proof of Theorem \ref{th5} ends.
\begin{figure*}
\begin{equation}\label{jointprocess}
	\begin{aligned}
		&\text{Prob}\left(\text{SINR}_k^i < \gamma_{\text{th}}, Q_k^i < Q_{\text{th}}\right), \forall k= 1, 2, \dots, K\\
		&\hspace{.5mm}=\text{Prob}\left(\frac{X_1}{\gamma_{\text{th}}}<Y_1<\widehat{Q}_{\text{th}}-X_1,\dots,\frac{X_K}{\gamma_{\text{th}}}< Y_K<\widehat{Q}_{\text{th}}-X_K\right) \\
		&\hspace{.5mm}=\int\cdots\int\text{Prob}\left(\left.\frac{X_1}{\gamma_{\text{th}}}<Y_1<\widehat{Q}_{\text{th}}-X_1,\dots,\frac{X_K}{\gamma_{\text{th}}}< Y_K<\widehat{Q}_{\text{th}}-X_K\right\vert X_1, \dots, X_K\right) f_{X_1,\dots,X_K}(x_1,\dots,x_K)dx_1\cdots dx_K\\
		&\overset{(a)}{=}\underbrace{\int_{0}^{\frac{\widehat{Q}_{\text{th}}}{\left(1+1/\gamma_{\text{th}}\right)}}\cdots\int_{0}^{\frac{\widehat{Q}_{\text{th}}}{\left(1+1/\gamma_{\text{th}}\right)}}}_{x_1,\dots,x_K}\left[F_{Y_1,\dots,Y_K}(\widehat{Q}_{\text{th}}-x_1,\dots,\widehat{Q}_{\text{th}}-x_K)-F_{Y_1,\dots,Y_K}\left(\frac{x_1}{\gamma_{\text{th}}},\dots,\frac{x_K}{\gamma_{\text{th}}}\right)\right]f_{X_1,\dots,X_K}(x_1,\dots,x_K)dx_1\cdots dx_K\\
		&\overset{(b)}{=}\underbrace{\int_{0}^{\frac{\widehat{Q}_{\text{th}}}{\left(1+1/\gamma_{\text{th}}\right)}}\cdots\int_{0}^{\frac{\widehat{Q}_{\text{th}}}{\left(1+1/\gamma_{\text{th}}\right)}}}_{x_1,\dots,x_K}\int_{r_1=0}^{\infty}\int_{r_2=0}^{\infty}\frac{r_2^{N-2}\exp{\left(-\frac{r_1+r_2}{2}\right)}}{2^{N}\Gamma(N-1)}\\
		&\enspace\enspace\enspace\times\prod_{k=1}^{K}\left[Q_{N-1}\left(\sqrt{\frac{\mu^2}{1-\mu^2}r_1},\sqrt{\frac{x_k}{\gamma_{\text{th}}}}\right)-Q_{N-1}\left(\sqrt{\frac{\mu^2}{1-\mu^2}r_1},\sqrt{\widehat{Q}_{\text{th}}-x_k}\right)\right]\\
		&\enspace\enspace\enspace\times\frac{1}{2}\exp\left(-\frac{x_k+\frac{\mu^2}{1-\mu^2}r_2}{2}\right)I_0\left(\sqrt{\frac{\mu^2}{1-\mu^2}}\sqrt{r_2x_k}\right)dr_1dr_2dx_1\cdots dx_K\\
		&\overset{(c)}{=}\int_{r_1=0}^{\infty}\int_{r_2=0}^{\infty}\frac{r_2^{N-2}\exp{\left(-\frac{r_1+r_2}{2}\right)}}{2^{N}\Gamma(N-1)}\prod_{k=1}^{K}\int_{x_k=0}^{\frac{\widehat{Q}_{\text{th}}}{\left(1+1/\gamma_{\text{th}}\right)}}\left[Q_{N-1}\left(\sqrt{\frac{\mu^2}{1-\mu^2}r_1},\sqrt{\frac{x_k}{\gamma_{\text{th}}}}\right)-Q_{N-1}\left(\sqrt{\frac{\mu^2}{1-\mu^2}r_1},\sqrt{\widehat{Q}_{\text{th}}-x_k}\right)\right]\\
		&\enspace\enspace\enspace\times\frac{1}{2}\exp\left(-\frac{x_k+\frac{\mu^2}{1-\mu^2}r_2}{2}\right)I_0\left(\sqrt{\frac{\mu^2}{1-\mu^2}}\sqrt{r_2x_k}\right)dx_kdr_1dr_2
	\end{aligned}
\end{equation}
\hrulefill
\end{figure*}

\renewcommand{\theequation}{G.\arabic{equation}}
\setcounter{equation}{0}
\section{Proof of Corollary 4}\label{c3}
First, for WDT, when $\mu$ is small, \eqref{mathcalC} is upper bounded by $ \frac{1-\mu^2}{\gamma_{\text{th}}}$. As a result, \eqref{FurtherWDT} can be lower bounded by
\begin{equation}\label{lowerWDT}
	\epsilon_{\text{WDT-SINR}}^i>\left[1-K\left(\frac{\mu^2}{\gamma_{\text{th}}+1}\right)^{N-1}-K\left(\frac{1-\mu^2}{\gamma_{\text{th}}}\right)^{N-1}\right]^{+}.
\end{equation}
Therefore, for large $N$, the second term and the third term of \eqref{lowerWDT}  are close to $0$, which means that the lower bound of WDT outage probability is close to $1$. Thus, the WDT outage probability  $\epsilon^i_{\text{WDT-SINR}}$ will be close to $1$ for large $N$.

As for WET, \eqref{ApproximateWET} can be lower bounded by (\ref{furtherWETapproximate}) (see top of next page), 
\begin{figure*}[]
	\begin{equation}\label{furtherWETapproximate}
		\begin{aligned}
			\epsilon^i_{\text{WET-EHP}} &\hspace{.5mm}\approx\left[1-K  \frac{\Gamma\left(N,\frac{\widehat{Q}_{\rm th}}{2}\right)}{\Gamma(N)}-K\frac{\mu^2\left(\frac{\widehat{Q}_{\rm th}}{2}\right)^{N}\exp\left({-\frac{\widehat{Q}_{\rm th}}{2}}\right)}{N!}\sum_{l=0}^{N-1}(1-\mu^2)^l{_1F_1\left(l+1;N+1;\frac{\mu^2\widehat{Q}_{\rm th}}{2}\right)}\right]^{+}\\
			&\overset{(a)}{>}1-K  \frac{\Gamma\left(N,\frac{\widehat{Q}_{\rm th}}{2}\right)}{\Gamma(N)}-K\frac{\mu^2\left(\frac{\widehat{Q}_{\rm th}}{2}\right)^{N}\exp\left({-\frac{\widehat{Q}_{\rm th}}{2}}(1-\mu^2)\right)}{N!}\sum_{l=0}^{N-1}(1-\mu^2)^l\\
			&\overset{(b)}{=}\left[1-K  \frac{\Gamma\left(N,\frac{\widehat{Q}_{\rm th}}{2}\right)}{\Gamma(N)}-K\frac{(1-(1-\mu^2)^N)\left(\frac{\widehat{Q}_{\rm th}}{2}\right)^{N}}{N!}\exp\left({-\frac{\widehat{Q}_{\rm th}}{2}}(1-\mu^2)\right)\right]^{+}\\
			&\overset{(c)}{\approx}\left[1-K  \frac{\Gamma\left(N,\frac{\widehat{Q}_{\rm th}}{2}\right)}{\Gamma(N)}\right]^{+}\\
			&\overset{(d)}{=}\left[1-K\exp\left(-\frac{\widehat{Q}_{\rm th}}{2}\right)\sum_{s=0}^{N}\frac{\left(\frac{\widehat{Q}_{\rm th}}{2}\right)^s}{s!}\right]^{+}
		\end{aligned}
	\end{equation}
	\hrulefill
\end{figure*}
where ($a$) accounts for the fact that \cite[eq. (13.6.1)]{olver2010nist}
\begin{align}
	{_1F_1\left(l;N+1;\frac{\mu^2\widehat{Q}_{\text{th}}}{2}\right)}&<{_1F_1\left(N+1;N+1;\frac{\mu^2\widehat{Q}_{\text{th}}}{2}\right)}\notag\\
	&=\exp\left(\frac{\mu^2\widehat{Q}_{\text{th}}}{2}\right),~\forall l,
\end{align}
($b$) makes some simplifications, $(c)$ accounts for the fact that when $\mu$ is small, $1-\mu^2 \approx 1$; hence the third term can be omitted, and ($d$) applies the result of \cite[eq. (8.4.10)]{olver2010nist}. As can be seen, the lower bound is a strictly decreasing function of $N$. Evidently, when $N$ is small, the lower bound in \eqref{furtherWETapproximate} is close to $1$. Thus, the WET outage probability $\epsilon^i_{\text{WET-EHP}}$ will be close to $1$ for small $N$.

	\bibliographystyle{ieeetr}
	%\bibliography{reference}
	
\end{document}